\documentclass{aa}  

\usepackage{graphicx}
\usepackage{subfigure}
\usepackage{txfonts}
\begin{document}

   \title{How external photoevaporation changes the inner disc's chemical composition
   }

   \author{N. Ndugu
          \inst{1,2}
          \and
          B. Bitsch\inst{3,4}
          \and 
          J. Lienert \inst{4}
          }

   \institute{Makerere University, P.O Box 7062, Kampala, Uganda\\
              \email{ndugunelson90@gmail.com}
                   \and
             HailNumbers Education Foundation, P.O Box 151972 , Kampala, Uganda
         \and
             University College Cork, College Rd, Cork T12 K8AF, Ireland
             \and
           Max Planck Institute for Astronomy (MPIA), Königstuhl 17, D-69117 Heidelberg  
             }

   \date{}

   \abstract{Stars mostly form in cluster environments, where neighbouring stars can have influence on the evolution of the newly formed protoplanetary discs. Besides gravitational interactions, external photoevaporation can also shape protoplanetary discs. Depending on the strength of external photoevaporation, discs can be destroyed within 1-2 Myrs, or more gradually, depending on whether the external photoevaporation field is stronger or weaker, respectively. We use the \texttt{chemcomp} code, which includes a viscous disc evolution model including pebble drift and evaporation to calculate the chemical composition of protoplanetary discs. We extend this code to include external photoevaporation following the FRIED grid. Before external photoevaporation becomes efficient, the disc follows a purely viscous disc evolution, where initially the C/O ratio in the inner disc decreases due to inward drifting and evaporating water ice pebbles. Over time, the C/O ratio increases again as water vapour is accreted onto the star and carbon-rich gas gradually migrates inward. However, once external photoevaporation starts to act, the outer disc begins to get dispersed. During this process, the inner disc's chemical evolution still follows the evolution of a purely viscous discs, because the majority of the pebbles have already drifted inwards on time-scales shorter than 1 Myr. At low viscosity, the inner disc's C/O ratio remains sub-solar until the disc is dispersed through external photoevaporation. At high viscosity, the inner disc's composition can reach super-solar values in C/O, because the water vapour is accreted onto the star faster and carbon rich gas from the outer disc can move inwards faster as well, as long as the disc can survive a few Myrs. In both cases, there is no visible difference in the chemical composition of the inner disc compared to a purely viscous model, due to the rapid inward drift of pebbles that sets the chemical composition of the disc. Our model thus predicts that the inner disc chemistry should be similar between discs that are subject to external photoevaporation and discs that are isolated and experience no external photoevaporation, in line with observations of protoplanetary discs with JWST.
   }

   \keywords{protoplanetary discs–photoevaporation–chemical evolution}

   \maketitle
%

\section{Introduction}

Planets form in protoplanetary discs, where planetary formation depends crucially on the different disc parameters like temperature, density and viscosity \citep[e.g.,][]{1972epcf.book.....S,1997ApJ...490..368C,2023A&A...679A..42S}. It is therefore essential to understand the evolution and structure of discs in the first place. Protoplanetary discs are composed primarily of hydrogen and helium, with a small fraction of dust grains that can grow into mm-cm sized pebbles \citep[e.g.,][]{1993prpl.conf.1031W,2008ARA&A..46...21B}, which are the first building blocks of planetesimals and planets \citep{2014prpl.conf..547J}.

These small mm-cm sized pebbles then start to drift inwards due to gas-drag effects \citep[e.g.,][]{weidenschillingAerodynamicsSolidBodies1977,brauerCoagulationFragmentationRadial2008}. As the pebbles move towards the inner disc regions, the disc environment heats up and these pebbles can eventually evaporate and release their volatiles into the gas phase of the disc \citep[e.g.,][]{2013A&A...552A.137R,2015ApJ...815..109P,2017MNRAS.469.3994B,schneiderHowDriftingEvaporating2021a,2022ApJ...936...40E,kalyaanEffectDustEvolution2023}.

The released volatile vapour is then accreted onto the central star on a viscous time-scale. Due to the velocity difference between the inward drifting pebbles and the gas \citep[e.g.,][]{brauerCoagulationFragmentationRadial2008}, a large enhancement of volatile vapour interior to evaporation fronts can be created \citep[e.g.,][]{schneiderHowDriftingEvaporating2021a,mahCloseinIceLines2023,kalyaanEffectDustEvolution2023}. This volatile enhancement of the inner disc has important consequences for the composition of growing planets \citep[e.g.,][]{schneiderHowDriftingEvaporating2021,2022A&A...665A.138B}. Additionally, this process also indicates that the disc's chemical composition is not static in time  \citep[e.g.,][]{molliereInterpretingAtmosphericComposition2022}.

The inward drift of pebbles that enriches the inner discs with volatiles can be influenced by pressure perturbations in the outer disc 
\citep[e.g.,][]{bitschDryWaterWorld2021,kalyaanLinkingOuterDisk2021,2024arXiv240209342L, 2024arXiv240606219M}, which stop the inward flux of pebbles. Theoretical predictions indicate that inner discs with bumps/gaps should be relatively volatile poor (especially in respect to water), while smooth discs should be water rich in the inner regions. Observationally, this trend seems to hold  \citep[e.g.,][]{banzattiJWSTRevealsExcess2023,gasmanMINDSAbundantWater2023, perottiWaterTerrestrialPlanetforming2023}, indicating that the disc structure is of uttermost importance in setting the inner disc chemistry.

Pressure perturbations in the outer disc can be caused by growing planets \citep[e.g.,][]{paardekooperDustFlowGas2006,pinillaTrappingDustParticles2012,lambrechtsSeparatingGasgiantIcegiant2014,bitschPebbleisolationMassScaling2018,ataieeHowMuchDoes2018}, ice line effects \citep[e.g.,][]{Pinilla_2017, 2021A&A...650A.185M}, or also by photoevaporation \citep[e.g.,][]{2008ApJ...688..398E, owenTheoryDiscPhotoevaporation2012a, picognaDispersalProtoplanetaryDiscs2019a, 2024arXiv240209342L}. Photoevaporation is a process where high-energy radiation from either the central star or external sources accelerates the particles of the disc and removes them from their bound orbits to allow an escape into space (see \cite{pascucciRoleDiskWinds2022} for a review). This process is fundamentally different compared to gap opening by giant planets, where part of the gas can still pass its orbit \citep[e.g.,][]{2008ApJ...685..560D, 2020A&A...643A.133B}, resulting in carbon-rich material that can eventually make it to the inner disc, in contrast to gaps opened by photoevaporation, where outer material can not make it to the inner disc any more, resulting in low C/O ratios \citep{2024arXiv240209342L}.

Stars are not born in isolation, but rather in star forming regions \citep[e.g.,][]{2003ARA&A..41...57L, 2008ApJ...675.1361F, 2020Natur.586..528W, 2024arXiv240302149G}. Consequently, their discs are subject to several factors that can complicate their evolution, like background heating \citep{nduguPlanetPopulationSynthesis2018}, stellar encounters \citep{2022MNRAS.512..861N}, and external photoevaporation \citep{2019MNRAS.485.3895H, pascucciRoleDiskWinds2022,2020Natur.586..528W}. External photoevaporation is caused by surrounding massive stars that emit ultraviolet (UV) radiation \citep[e.g.,][]{1998AJ....116..322H,1999AJ....118.2350H}.

In our past study, we explored how internal photoevaporation \citep{2008ApJ...688..398E,2019MNRAS.487..691P,2020MNRAS.492.1279S} influences the disc's composition \citep{2024arXiv240209342L}. Here, we want to expand on this approach by studying the effects of external photoevaporation on the structure, evolution and composition of protoplanetary discs. External photoevaporation mainly affects the outer disc regions \citep{2020MNRAS.492.1279S} by slowly eating them away. However, the outer disc regions harbour the pebble reservoir of the disc, and its evaporation might hinder the efficient inward drift of pebbles and might thus prevent the formation of planets \citep[e.g.,][]{2023MNRAS.526.4315H, 2023MNRAS.522.1939Q}, if photoevaporation acts fast enough. In order to model external evaporation, we make use of the FRIED grid \citep{2023MNRAS.526.4315H}, which gives simple recipes to include external photoevaporation into viscous disc evolution models without the necessity to run detailed simulations.

To achieve our objectives, we employ the \texttt{chemcomp} tool, a one-dimensional, semi-analytical model that simulates the evolution of protoplanetary discs and the initial stages of planet formation \citep{schneiderHowDriftingEvaporating2021}. This tool integrates the effects of viscous evolution, pebble drift and evaporation, and allows to study the chemical composition of the disc. By examining the variations in the gas surface densities, the ratios of key elements such as carbon, oxygen, and nitrogen over time, we aim to understand the potential impact of external photoevaporation on the disc's composition and consequently on planet formation and the implications for their eventual composition.

Our study is organized as follows: Section~\ref{sec:Methods} outlines the methods used, including the \texttt{chemcomp} model and the parameters employed in our simulations. We furthermore show how we include external photoevaporation. Section~\ref{sec:Results} presents our results, highlighting the effects of external photoevaporation on the disc's structure and chemical composition. In Section~\ref{sec:Compar}, we compare our results to previous studies of photoevaporation and to disc observations, setting the stage for future research perspectives, and discuss the broader implications of our findings for the process of planet formation and the characteristics of the resulting planetary systems before giving a summary of our findings in section~\ref{sec:summary}.

\section{Methods}\label{sec:Methods}

In order to study the effects of external photoevaporation on the chemical composition of inner discs, we extend the \texttt{chemcomp} code \citep{schneiderHowDriftingEvaporating2021}. \texttt{chemcomp} is a one-dimensional semi-analytical model that simulates the viscous evolution of protoplanetary discs and how planets can form in them. In particular, the viscous evolution is modelled using the $\alpha$-viscosity \citep{lynden-bellEvolutionViscousDiscs1974a, shakuraBlackHolesBinary1973} approach. Furthermore, the code models the growth and inward drift of particles using the method of \citet{birnstielSimpleModelEvolution2012}, while simultaneously tracking the composition of the growing grains and how their evaporation at evaporation fronts changes the volatile composition of the gas and solids \citep{schneiderHowDriftingEvaporating2021}. The initial chemical abundances are listed in appendix~\ref{app:parameters} and are the same as in our previous works \citep{schneiderHowDriftingEvaporating2021, schneiderHowDriftingEvaporating2021a, bitschHowDriftingEvaporating2022}. We do not include changes of the chemical compositions due to chemical reactions, as the chemical reaction time-scale is normally longer than the pebble drift time-scale \citep[e.g.,][]{boothPlanetformingMaterialProtoplanetary2019}. \texttt{chemcomp} also includes a planet formation model, which we will not use in this work, as we will focus on the disc's evolution. The base version of the code (without the herein discussed additions of external photoevaporation) is publicly available \citep{2024arXiv240115686S}. In the following, we recap the important ingredients to our model, more details can be found in \citet{schneiderHowDriftingEvaporating2021}.

 \begin{figure*}[h]
    \centering
    \begin{minipage}{\textwidth}
        \subfigure{\includegraphics[width=0.5\textwidth]{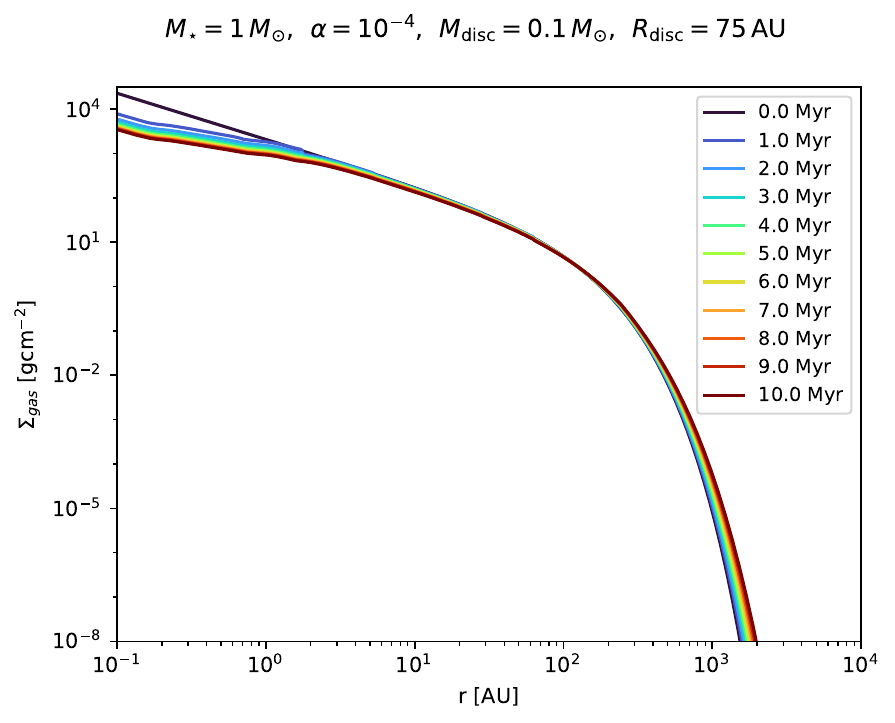}}
        \subfigure{\includegraphics[width=0.5\textwidth]{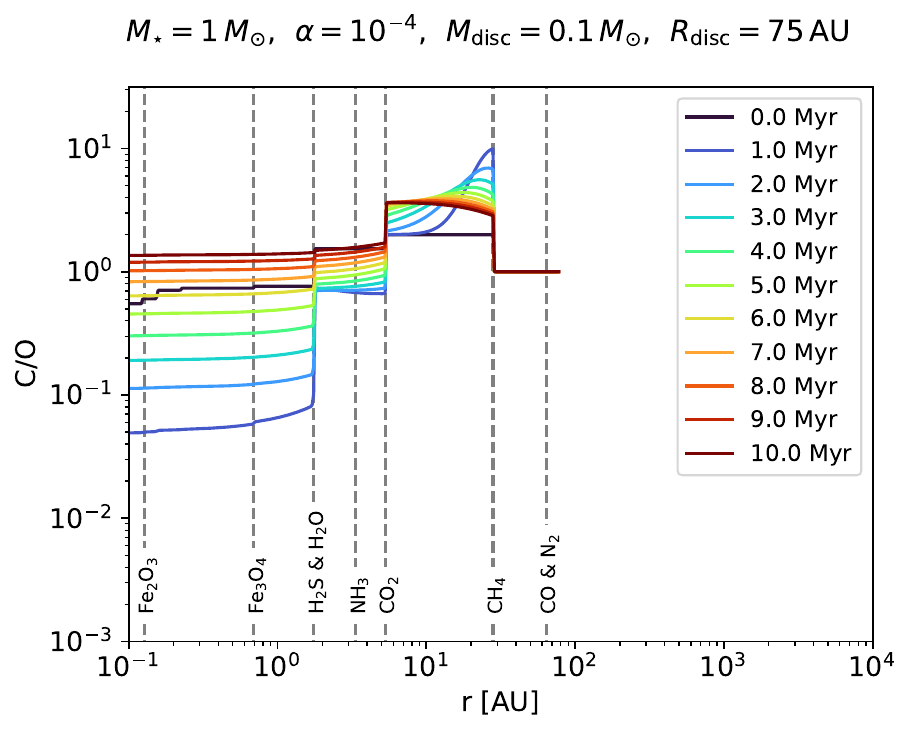}}
    \end{minipage}
    \caption{Nominal gas surface density (left panel) and the nominal carbon to oxygen fraction (right panel) as a function of disc radius for a viscous disc evolution, as given in equation~\ref{eq:viscous_disc_equation}. The simulations here represent the parameters outlined in Table~\ref{table:sim_par}, matching our standard simulations. The time evolution is shown in colour - from black, which corresponds to 0 Myr, to dark red, which corresponds to 10 Myr.}
    \label{fig:sig_gas}
\end{figure*}

\subsection{Viscous evolution}

The viscosity is given by
\begin{equation}
    \nu = \alpha \frac{c_s^2}{\Omega_{\text{K}}},
\end{equation}
where $\alpha$ is a dimensionless factor describing the strength of the turbulence, $c_s$ is the isothermal sound speed and $\Omega_{\text{K}} = \sqrt{\frac{GM_{\star}}{r^3}}$ is the Keplerian angular frequency. Here, $r$ is the orbital distance, $G$ the gravitational constant and $M_\star$ the stellar mass, which we set to be the solar mass throughout this work. The speed of sound can be linked to the mid-plane temperature of the disc. The temperature is calculated from an equilibrium between viscous and stellar heating with radiative cooling. In our model, it does not evolve in time for simplicity and is thus calculated only at the initialization of the simulation. As viscous heating scales with the disc's surface density, discs with higher surface density have higher temperature, resulting in evaporation fronts that are located further away from the star within the inner region where viscous heating dominates. \\

The time evolution of the gas surface density in our disc is given by the viscous disc equation, which can be derived from the conservation of mass and angular momentum \citep{pringleAccretionDiscsAstrophysics1981,armitageAstrophysicsPlanetFormation2013},
\begin{equation} \label{eq:viscous_disc_equation}
    \frac{\partial \Sigma_{\text{gas,Y}}}{\partial t} - \frac{3}{r} \frac{\partial}{\partial r} \left[ \sqrt{r} \frac{\partial}{\partial r} \left( \sqrt{r} \nu \Sigma_{\text{gas,Y}} \right) \right] = \dot{\Sigma}_{\text{Y}} - \dot{\Sigma}_{\text{ext}},
\end{equation}
where $\Sigma_{\rm gas, Y}$ is the gas surface density of the molecular species Y and $\dot{\Sigma}_{\text{Y}}$ is the source term of the molecular species Y, as given in table \ref{table:molecular_species}. It originates from the evaporation and condensation of pebbles and is given by
\begin{equation}
\label{eq:evap}
    \dot{\Sigma}_{\text{Y}} = \left\{
        \begin{matrix}
            \dot{\Sigma}_{\text{Y}}^{\text{evap}} \quad r <    r_{\text{ice,Y}} \\
            \dot{\Sigma}_{\text{Y}}^{\text{cond}} \quad r \geq r_{\text{ice,Y}}.
        \end{matrix}
    \right.
\end{equation}
Here, $\dot{\Sigma}_{\text{Y}}^{\text{evap}}$ and $\dot{\Sigma}_{\text{Y}}^{\text{cond}}$ are the evaporation and condensation source terms of species Y, originating from the evaporation and condensation of volatiles of species Y at the respective ice line $r_{\text{ice,Y}}$. The term $\dot{\Sigma}_{\text{ext}}$ on the right hand side of the equation~\ref{eq:evap} describes the mass loss through external photoevaporation, discussed below.

In Figure~\ref{fig:sig_gas} we show the evolution of gas surface density and its corresponding carbon to oxygen ratio in a viscous disc without external photoevaporation. The parameters used for this simulations are shown in table~\ref{table:sim_par}. The low viscosity results in a very slow evolution of the disc's surface density. In appendix~\ref{app:chemistry} we show the detailed C/H, O/H and N/H evolution of this disc.

\begin{table}
    \centering
	\begin{tabular}{cc}
	\hline
        Simulation parameter                    & Value             \\ \hline \hline
        Stellar mass $M_{\star}$                & $1 \, M_{\odot}$  \\
        Stellar luminosity $L_{\star}$          & $1 \, L_{\odot}$  \\
        Viscous parameter $\alpha$              & $10^{-4}$         \\
        Initial disc mass $M_{\text{disc}}$     & $0.1 \, M_{\odot}$\\        
        Initial disc radius $R_{\text{disc}}$   & $75 \, \text{AU}$	\\
        Initial dust-to-gas ratio               & $2 \%$            \\
        UV field strength                       & $10^{4} \mathbf{G}_0$ \\ \hline
	\end{tabular}
    \caption{List of parameters used for our standard simulations.}
    \label{table:sim_par}
\end{table}

\subsection{Disc lifetime and external photoevaporation}

When a viscous disc is subject to external photoevaporation, an additional term $\dot{\Sigma}_{\text{ext}}$, describing the photoevaporative gas surface density loss rate, is subtracted from the right-hand-side of the disc's viscous equation (e.q.~\ref{eq:viscous_disc_equation}). In this study we focus on external photoevaporation \citep{2023MNRAS.526.4315H} and study how this affects the chemical composition of inner discs in order to be complementary to our previous study, which focused on the effects of internal photoevaporation by X-rays on the disc's chemical composition \citep{2024arXiv240209342L}. Our aim is to understand how the different evaporation mechanisms influence the composition of inner discs.

In order to simulate external photoevaporation, we make use of the publicly available FRIED grid \citep{2023MNRAS.526.4315H}. The FRIED grid originates from extensive simulations that study how external photoevaporation affects the mass loss rates of discs. The grid spans a wider range of model parameters, including disc size (1-500 AU), disc mass, strength of the ambient UV field ($10-10^4 \text{G}_0$) and stellar masses ($0.05 - 1.9 \text{M}_{\odot}$). We use linear interpolation to calculate the mass loss rates for our specific disc properties. Within the range of the disc properties analysed, the estimated mass loss rates span from $10^{-8}$ to $10^{-5}$ solar masses per Myr.

As the disc is dispersed away from the outer regions, one needs to define the outer edge of the disc. This is given by the transition from the optically thick to the optically thin medium, corresponding to the location of the maximum mass loss rate given by the FRIED grid \citep{2020MNRAS.492.1279S}. We assume a uniform mass loss rate of the outer 10\% of the disc's radius, in line with \citet{2019MNRAS.485.3895H}. The external photoevaporation rate is then given by

\begin{equation}
\dot{\Sigma}_{\text{ext}} = 
\begin{cases} 
0 & \text{for } r < \beta_{\text{ext}}R_{\text{edge}} \\
\frac{\dot{M}_{\text{ext}}}{\pi(R^2_{\text{edge}}-\beta^2_{\text{ext}}R^2_{\text{edge}})} & \text{for } r \geq \beta_{\text{ext}}R_{\text{edge}}
\end{cases} \ ,
\label{evap_eqn}
\end{equation}
where $\beta_{\text{ext}} = 0.9$, $R_{\text{edge}}$ is the outer edge of the gas disc and $\dot{M}_{\text{ext}}$ is the mass loss rate given by the FRIED grid. A smoothing term of $\beta_{\text{ext}}R_{\text{edge}}$ was applied to avoid numerical difficulties. Throughout this paper, we employ a UV field strength of \(10^{4} \mathbf{G}_0\), aiming at exploring the effects of an extremely effective photoevaporation field on various disc parameters and their interplay. If the UV field strength is very low, we expect no changes in the disc's evolution \citep{2023MNRAS.522.1939Q}, as we also show in subsection~\ref{app:photo}.

\subsection{Pebble drift and evaporation}

Protoplanetary discs are composed of gas and dust grains. Initially, we assume that the dust grains in the disc are of micro-meter size, similar as in the ISM. These particles can then grow to mm-cm sized pebbles due to coagulation \citep[e.g.,][]{brauerCoagulationFragmentationRadial2008} or condensation \citep[e.g.,][]{2013A&A...552A.137R, schneiderHowDriftingEvaporating2021a, mousisSituExplorationAtmospheres2022,2022ApJ...936...40E} at evaporation fronts. The maximal grain size is restrained by drift and fragmentation of colliding grains. The outcome of grain collisions depends on the collision properties \citep[e.g.,][]{2010A&A...513A..56G}, where especially the velocities with which the grains collide determine the outcome. In simulations, normally a so-called fragmentation velocity is set that gives the maximal velocities that grains can collide with to still grow. Larger velocities would result in fragmentation. Typically, these fragmentation velocities are between 1-10 $\rm ms^{-1}$ \citep[e.g.,][]{gundlachSTICKINESSMICROMETERSIZEDWATERICE2014,musiolikContactsWaterIce2019}. Here, we use a constant fragmentation velocity of 5 $\rm ms^{-1}$.

The radial motion of the dust grains within the disc is characterized by the radial dust velocity \(u_{\text{Z}}\), as described by the following equation:
\begin{equation}
    u_{\text{Z}} = \frac{1}{1 + \text{St}^2} u_{\text{gas}} - \frac{2}{\text{St}^{-1} + \text{St}} \Delta v,
\end{equation}
where \(\text{St}\) denotes the Stokes number of a particle (which is proportional to the particle's size), and \(u_{\text{gas}}\) and \(\Delta v\) represent the gas velocity and its azimuthal speed difference, respectively. The latter is defined as:
\begin{equation}
    \Delta v = v_{\text{K}} - v_{\varphi} = - \frac{1}{2} \frac{\text{d} \ln P}{\text{d} \ln r} \left( \frac{H_{\text{gas}}}{r} \right)^2 v_{\text{K}},
\end{equation}
where \(v_{\text{K}} = \Omega_{\text{K}} r\) is the Keplerian velocity, \(P\) is the gas pressure, and \(H_{\text{gas}}\) is the gas scale height.

We adopt here the simple two-population approach \citep{birnstielSimpleModelEvolution2012}, which divides the dust population into two size regimes, related to small and large grains. The dust is then advected with the average velocity of the two size regimes combined. Consequently, pressure perturbations that can stop the inward flux of particles will be perfect pebble traps in this model, whereas in reality, smaller dust grains can diffuse through the pressure bumps \citep[e.g.,][]{weberCharacterizingVariableDust2018,2019ApJ...885...91D,2023A&A...670L...5S}, depending on the pebble size and disc turbulence levels \citep[e.g.,][]{ataieeHowMuchDoes2018,bitschPebbleisolationMassScaling2018}--implementing this requires more detailed models. However, here we study the cases of smooth discs without pressure perturbations, justifying the simple approach.

Consequently, as the pebbles start growing, they decouple from the gas and begin to drift inwards with velocities much larger than the gas velocity \citep[e.g.,][]{weidenschillingAerodynamicsSolidBodies1977,brauerCoagulationFragmentationRadial2008}, resulting in a decay of the pebble flux within 1-2 Myr (see appendix~\ref{app:flux}). As the pebbles drift inwards, they move towards hotter disc regions, where they can eventually evaporate, as we discuss below.

\subsection{Chemical composition}

We follow the assumption that the initial composition of the disc mirrors the composition of the host star, where we use solar compositions as starting values (see appendix~\ref{app:parameters}). Towards the inner disc regions, the temperature increase allows the evaporation of first volatiles (e.g., H$_2$O, CO$_2$, NH$_3$, CH$_4$, CO) and then more refractory components.

In our approach, we allow that molecules of species Y, of a given composition, can exist either as solids (or ices) or in gaseous form, depending on their evaporation/condensation temperature. The mid-plane temperature that equals the molecules' evaporation temperature marks the evaporation line for that species (see table~\ref{table:molecular_species}). The evaporation of inward drifting pebbles is supposed to happen within 0.001~AU inwards of the evaporation front, while the re-condensation of outwards diffusing vapour depends on the efficiency that vapour sticking on already existing grains (see \citet{schneiderHowDriftingEvaporating2021} for more details). This effect can cause spikes in the pebble surface density distribution at ice lines, which depend also on the disc's viscosity \citep[e.g.,][]{schneiderHowDriftingEvaporating2021a,2022ApJ...936...42E}, and determines the outward transport of vapour.

As the pebbles drift inward and evaporate, they can enrich the inner disc with volatile vapour due to the velocity differences between fast inward moving pebbles and the slow gas transport. This effect can lead to large enrichment to super-solar values of volatiles in the inner disc region \citep[e.g.,][]{schneiderHowDriftingEvaporating2021a,bitschEnrichingInnerDiscs2023,kalyaanEffectDustEvolution2023,mahCloseinIceLines2023}. Generally, low viscosity discs are enriched to larger values \citep[e.g.,][]{mahCloseinIceLines2023,bitschEnrichingInnerDiscs2023} due to the larger pebbles that allow a faster inward motion and the slower gas transport, which is determined by the disc's viscosity.

We calculate the elemental ration between two species \(\text{X}_1\) and \(\text{X}_2\) as follows,
\begin{equation}
    \text{X}_1 / \text{X}_2 = \frac{m_{\text{X}_1}}{m_{\text{X}_2}} \frac{\mu_{\text{X}_2}}{\mu_{\text{X}_1}},
\end{equation}
where \(m_{\text{X}_1}\) and \(m_{\text{X}_2}\) are the mass fractions, and \(\mu_{\text{X}_1}\) and \(\mu_{\text{X}_2}\) are the atomic masses of the respective elements. This approach is used to calculate the C/O ratio and is the same as in our previous works.


\subsection{Initial parameters}

The stellar parameters in our simulations are fixed to solar mass and solar luminosity, for simplicity. As we want to study the importance of the disc parameters on external photoevaporation, we vary those. In particular, we use $\alpha$ parameters with values of $10^{-4}, 3\times 10^{-4}, 10^{-3}$, and disc masses of either 0.01 or 0.1 solar masses, with radii of either 75 or 150 AU. In addition, we test three different starting times for external photoevaporation (0, 1, and 3 Myr after initialisation). For all simulations, we use a dust-to-gas ratio of 2\%. Our simulation set-up thus follows the set-up of \citet{2024arXiv240209342L}, allowing an easy comparison to the simulations with internal X-ray driven photoevaporation.

For the initialization of our simulations, we use the analytical
solution (without the photoevaporation source term) found by \cite{lynden-bellEvolutionViscousDiscs1974a} with an exponential decay at large distances. This is the same starting condition as in our previous works.

\section{Results}\label{sec:Results}

   \begin{figure*}[h]
   \centering
   \includegraphics[width=\textwidth]{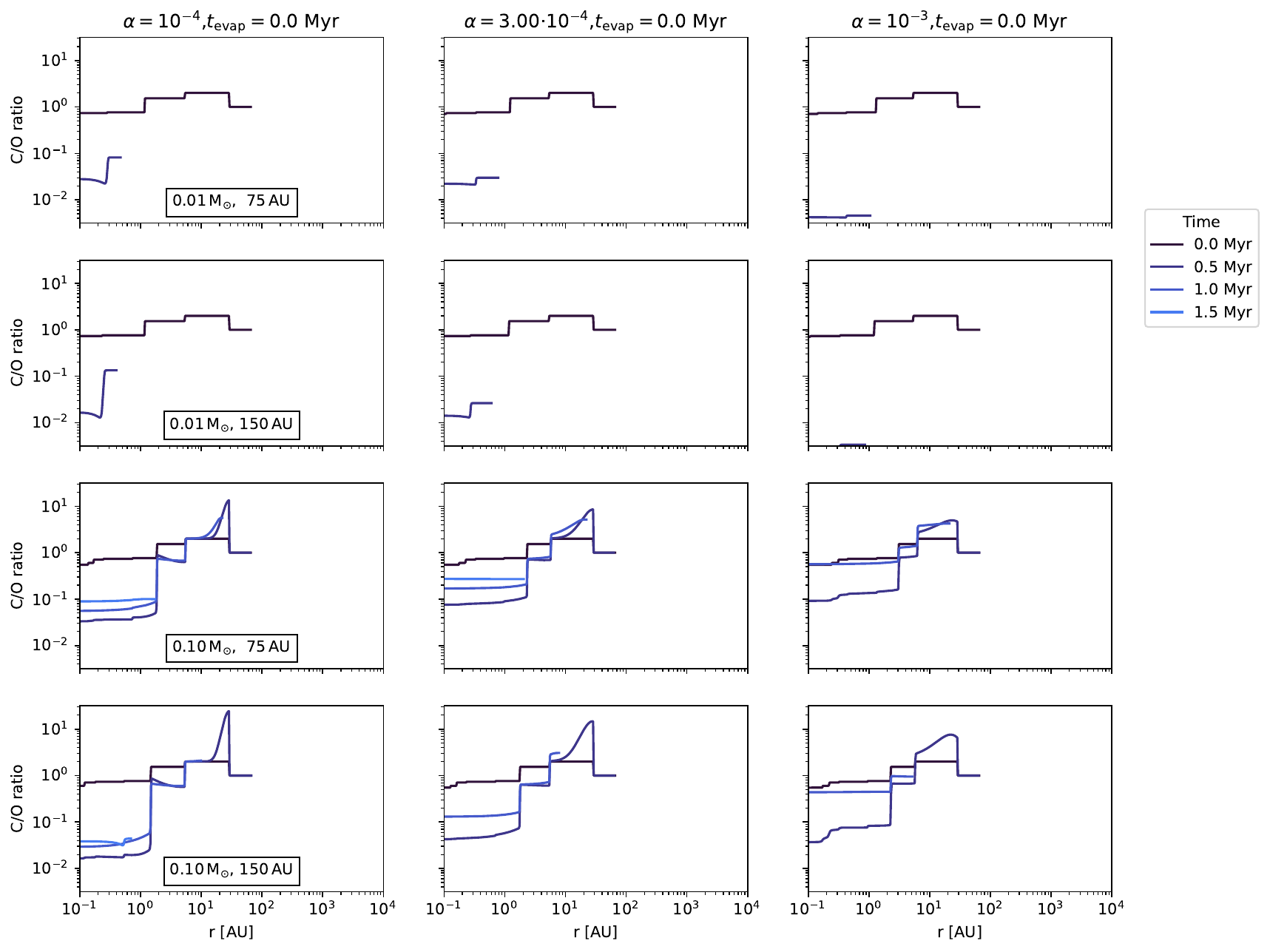}
   \caption{Carbon to Oxygen ratio of the protoplanetary discs as a function of time, corresponding to the disc profile shown in Figure~\ref{tevap0} with an onset of external photoevaporation at t=0.}
              \label{cotevap0}%
    \end{figure*}

\begin{figure*}[h]
   \centering
   \includegraphics[width=\textwidth]{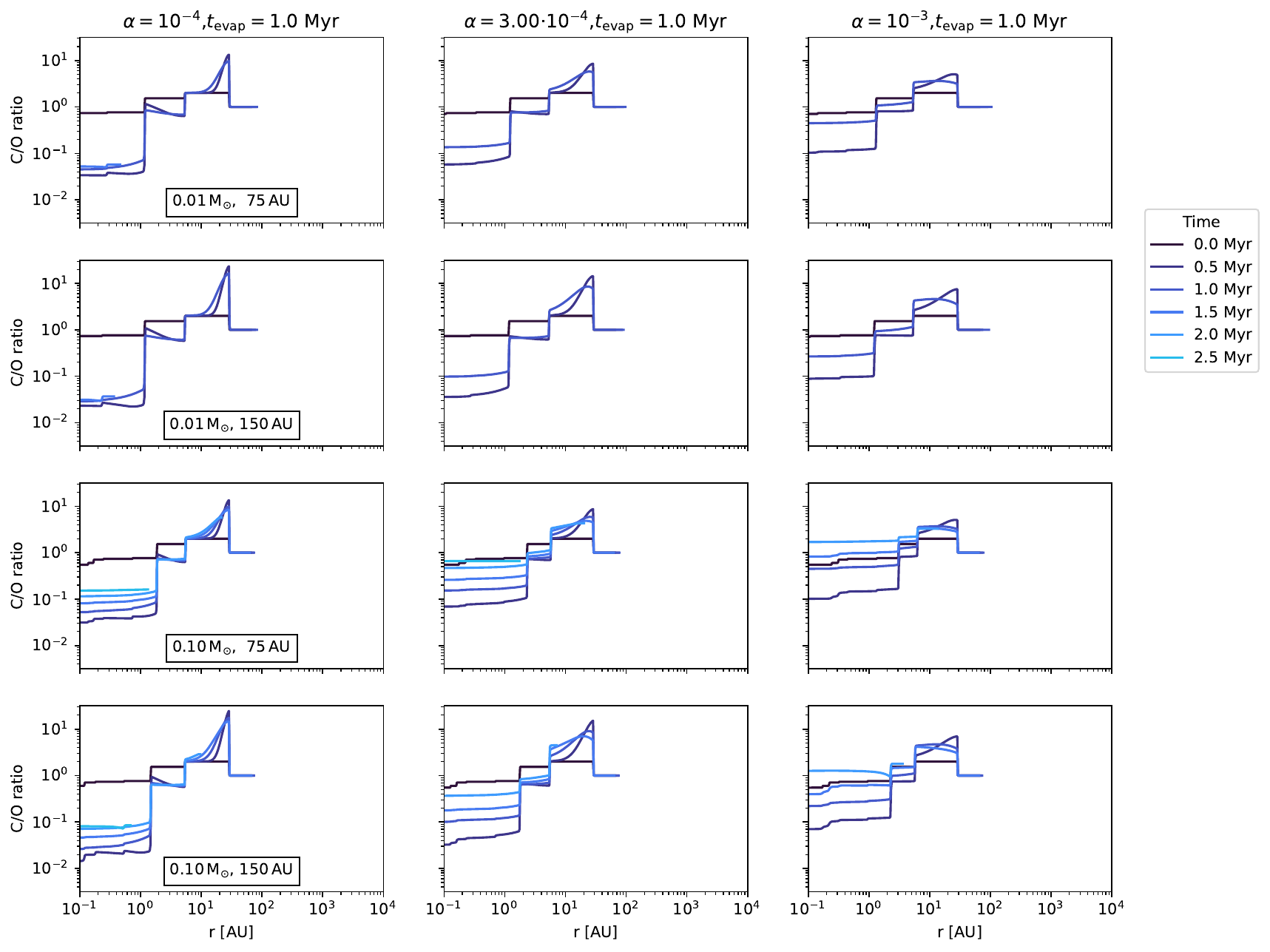}
   \caption{Carbon to oxygen ratio of the protoplanetary discs as a function of time and radius, corresponding to the disc profiles shown in Figure~\ref{tevap1} with an onset of external photoevaporation at t=1 Myr.}
              \label{cotevap1}%
    \end{figure*}

  \begin{figure*}[h]
   \centering
   \includegraphics[width=\textwidth]{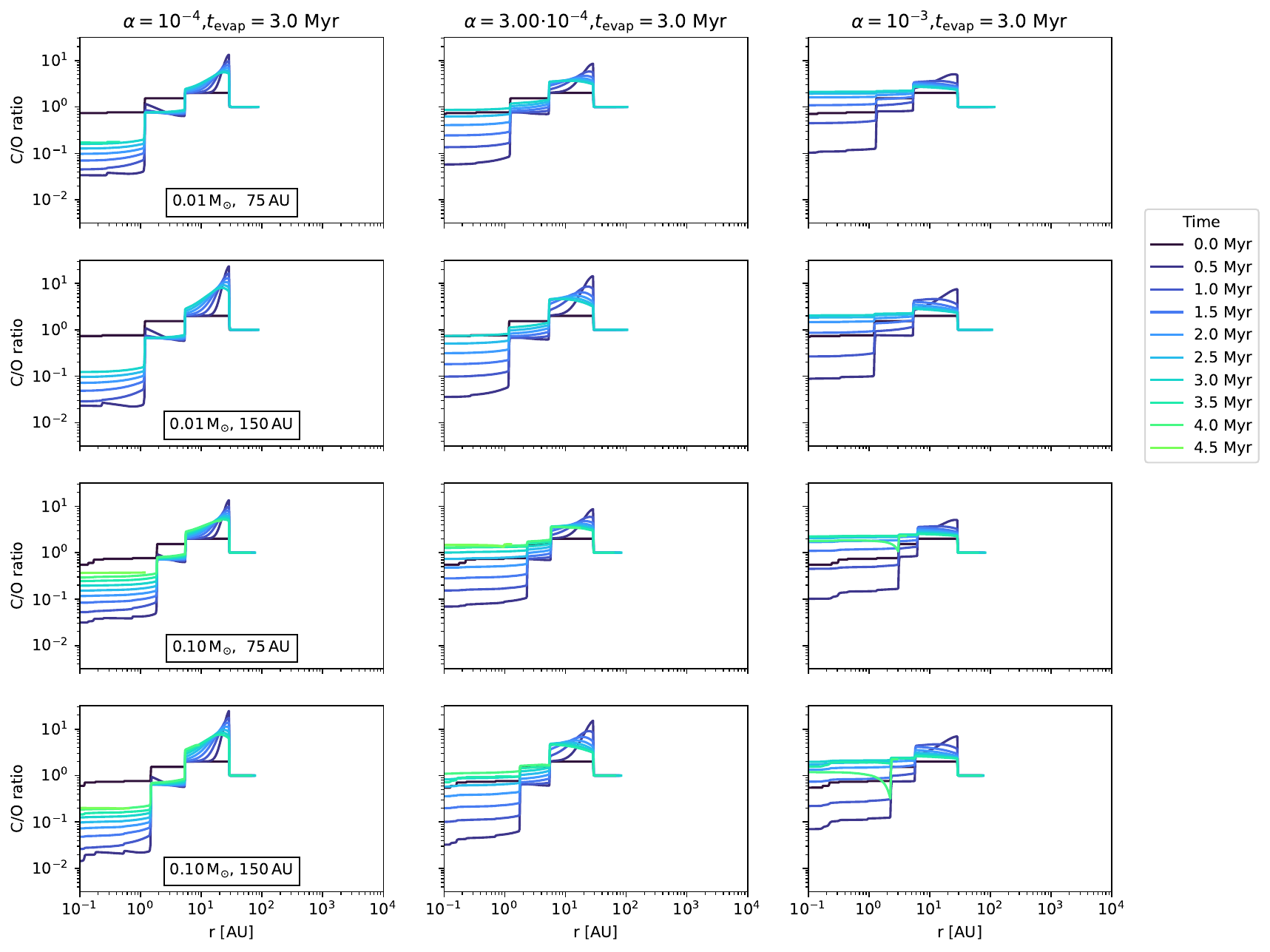}
   \caption{Carbon to Oxygen ratio of the protoplanetary discs as a function of time, corresponding to the disc profile shown in Figure~\ref{tevap3} with an onset of external photoevaporation at t=3 Myr.}
              \label{cotevap3}%
    \end{figure*}

    \begin{figure*}[h]
    \centering
    \begin{minipage}{\textwidth}
        \subfigure{\includegraphics[width=0.33\textwidth]{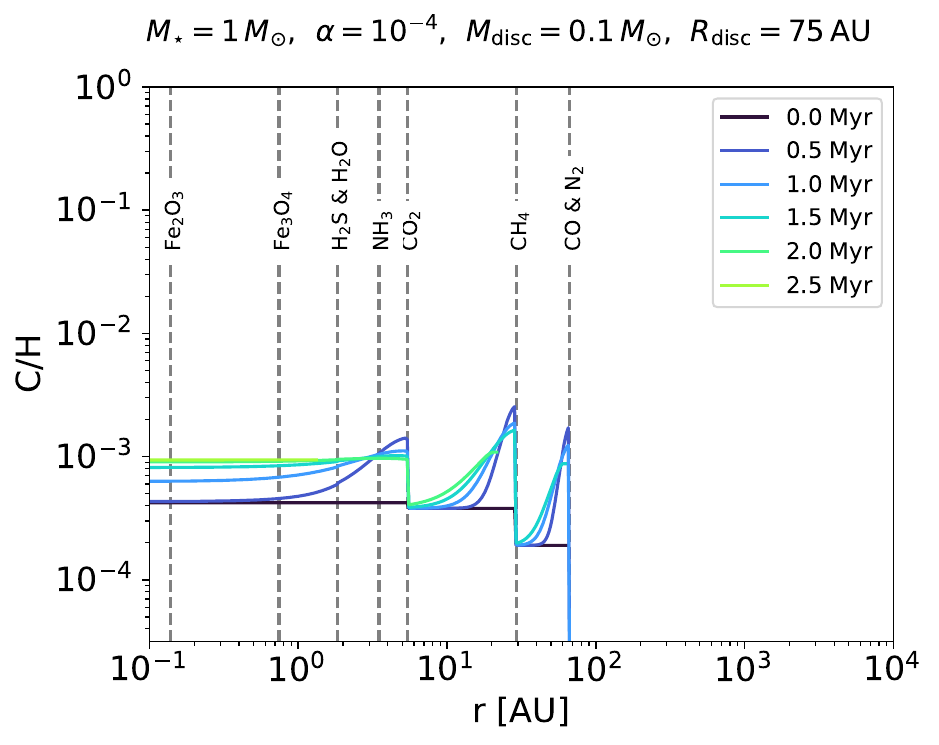}}
        \subfigure{\includegraphics[width=0.33\textwidth]{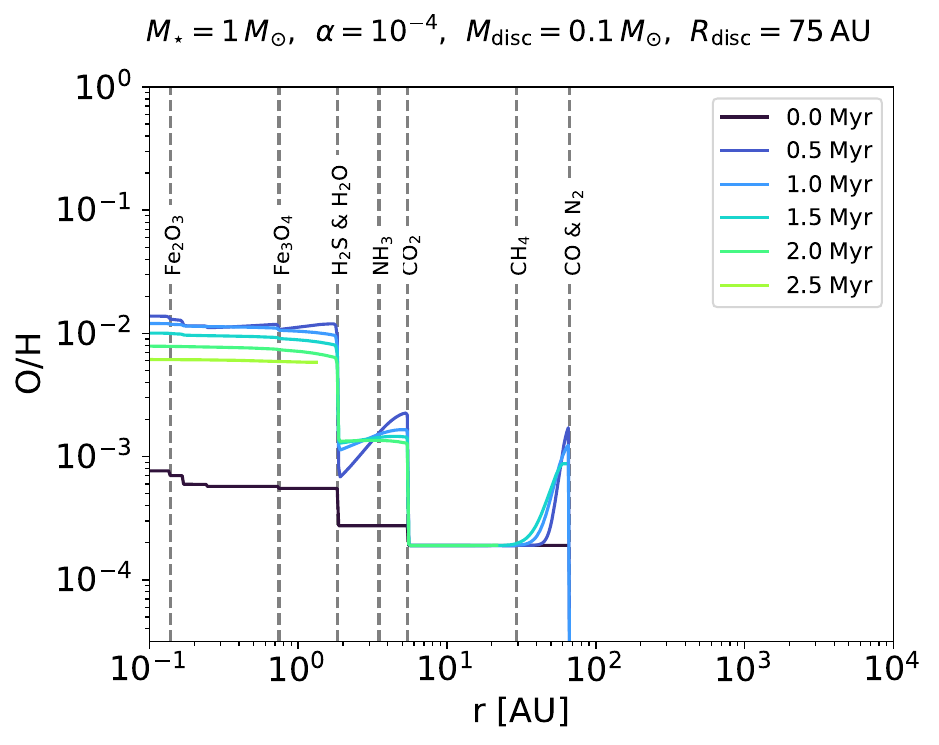}}
        \subfigure{\includegraphics[width=0.33\textwidth]{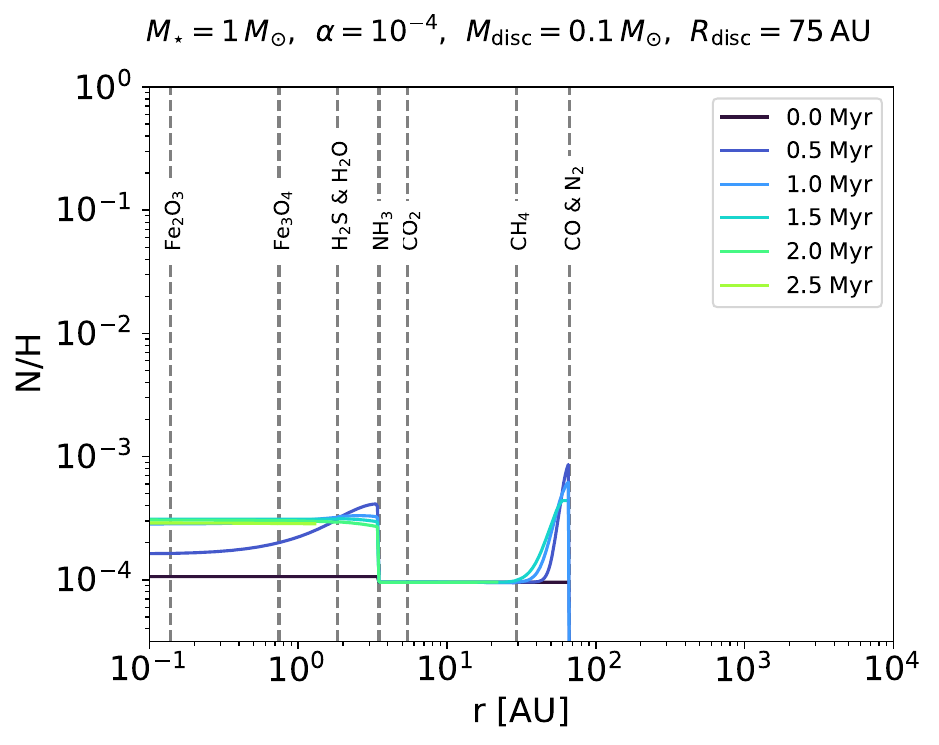}}
    \end{minipage}
    \caption{Different element ratios in the gas phase as a function of disc radius and time for a viscous disc with external photoevaporation begining after 1 Myr. The element ratios feature disc with viscosity of 1e-4, disc mass of 0.1 $M_{\odot}$, and disc radius of 75~au. Left: C/H, middle: O/H, right: N/H. The color coding shows the time evolution.}
    \label{fig:CH_OH_NH_with_ph}
\end{figure*}

\begin{figure*}[h]
    \centering
    \begin{minipage}{\textwidth}
        \subfigure{\includegraphics[width=0.33\textwidth]{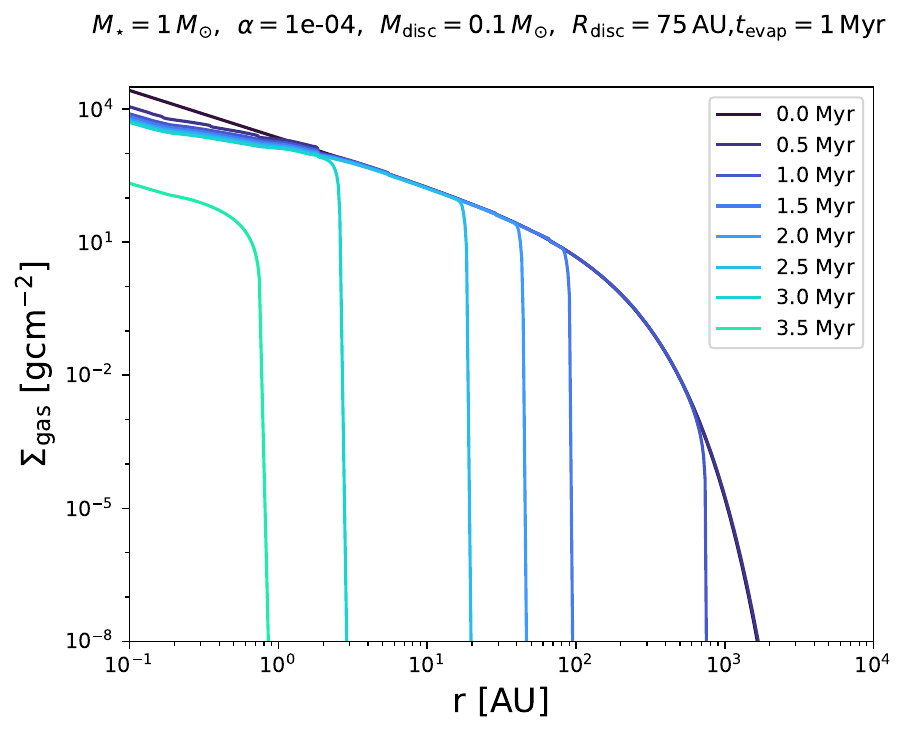}}
        \subfigure{\includegraphics[width=0.33\textwidth]{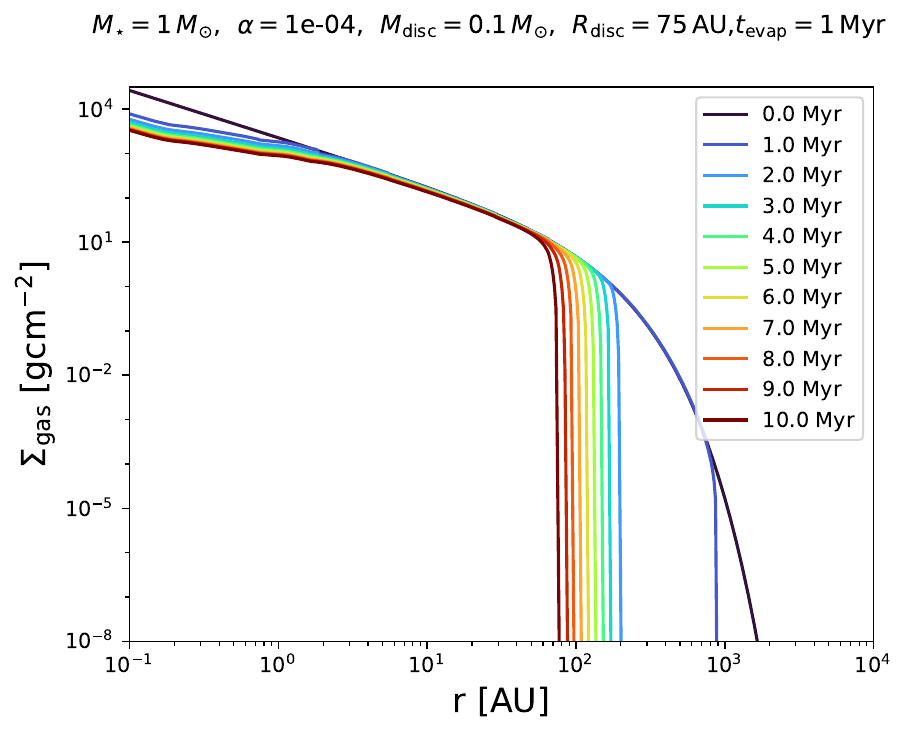}}
        \subfigure{\includegraphics[width=0.33\textwidth]{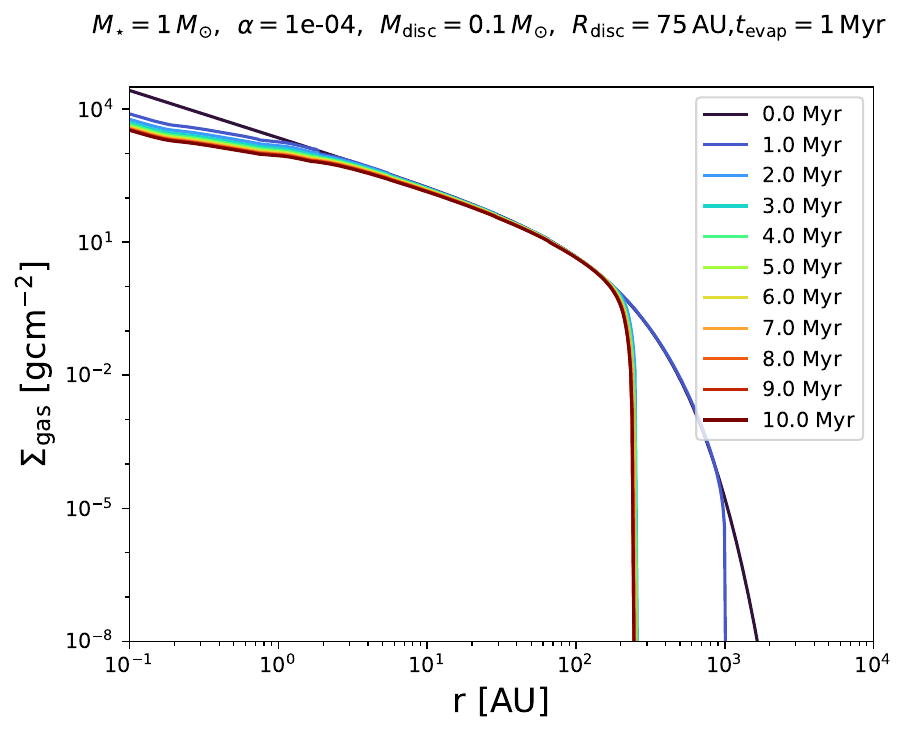}}
    \end{minipage}    
    \begin{minipage}{\textwidth}
        \subfigure{\includegraphics[width=0.33\textwidth]{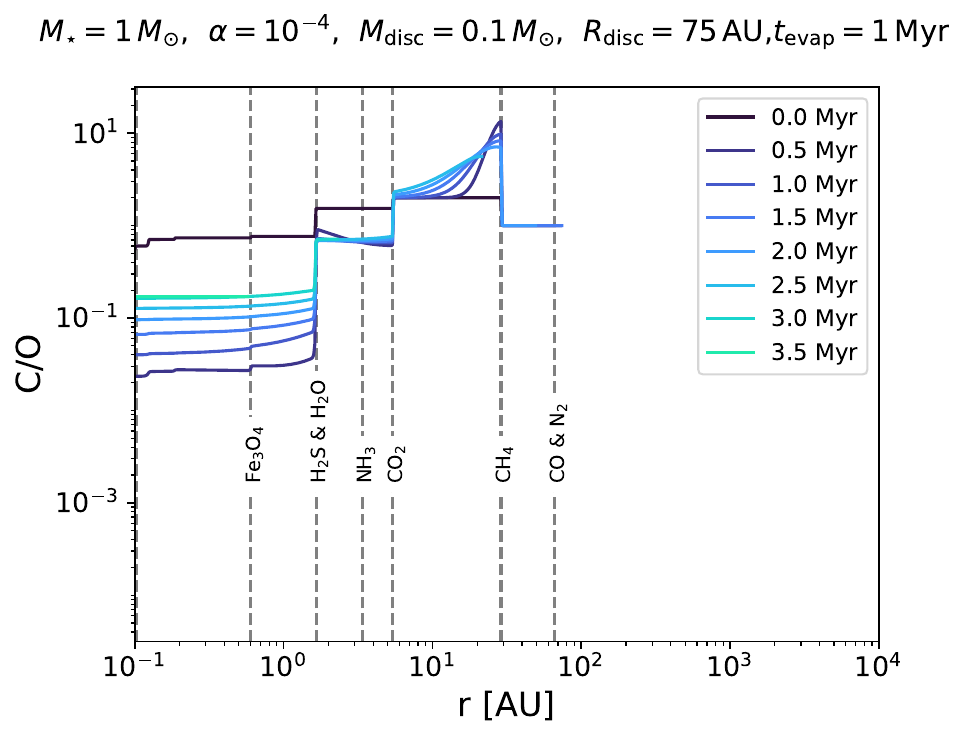}}
        \subfigure{\includegraphics[width=0.33\textwidth]{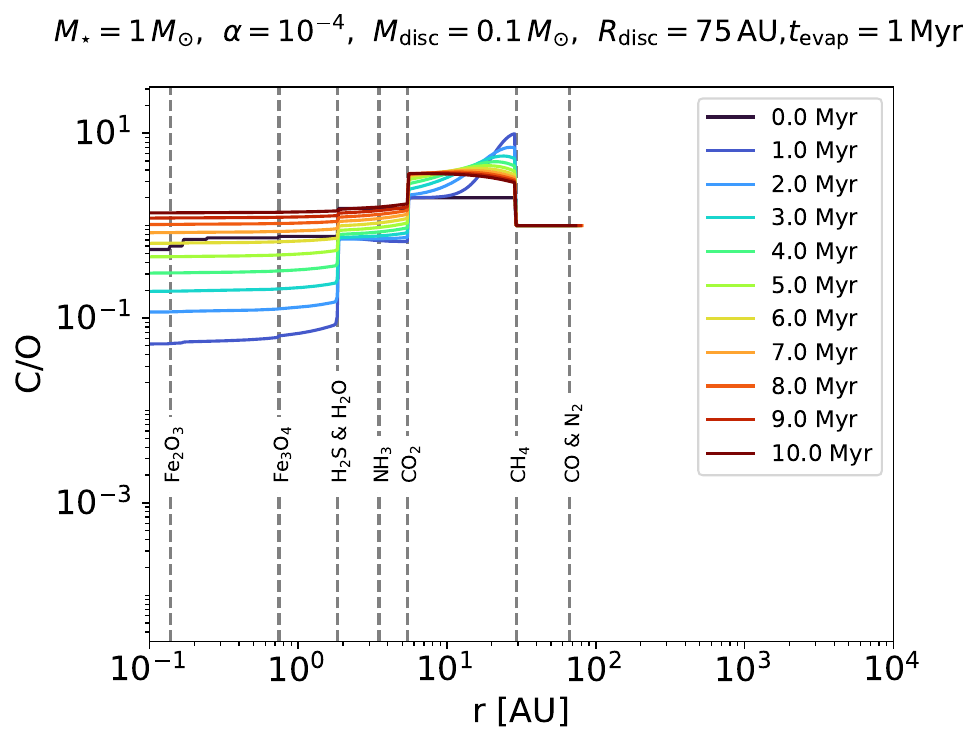}}
        \subfigure{\includegraphics[width=0.33\textwidth]{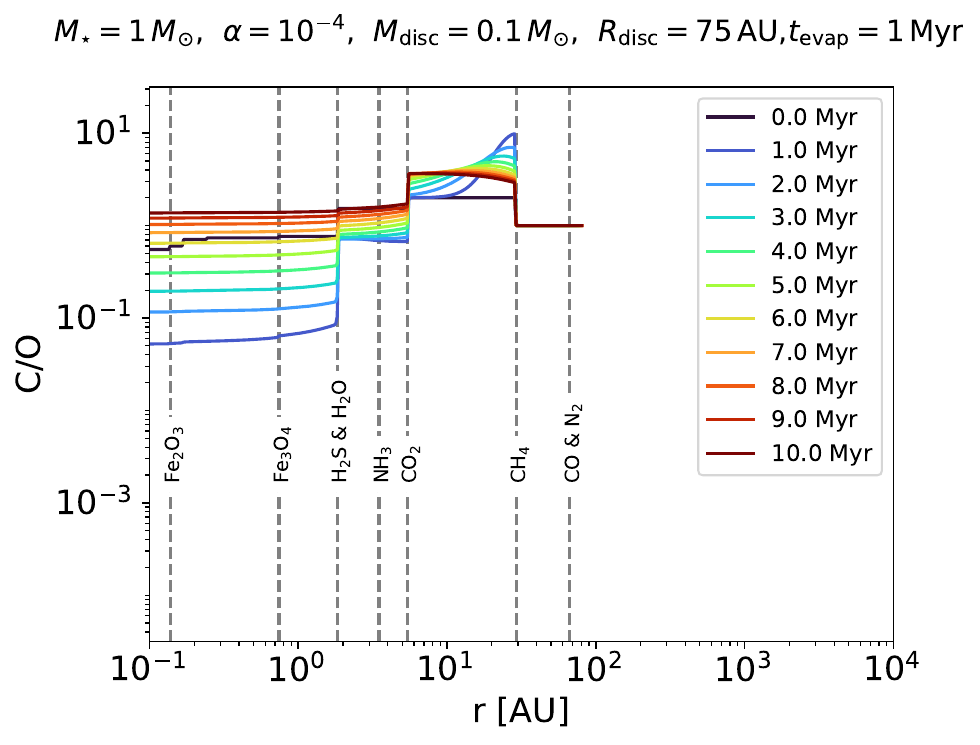}}
    \end{minipage}
    
    \caption{Gas surface density (top) and C/O ratio (bottom) for external photoevaporation strengths of \(10^{3} \mathbf{G}_0\) (left), \(10^{2} \mathbf{G}_0\) (center), and \(10 \mathbf{G}_0\) (right), for discs with external photoevaporation beginning at 1 Myr.}
    \label{fig:CO_ratio_with_1000G0-nominal}
\end{figure*}

   \begin{figure*}[h]
   \centering
   \includegraphics[width=\textwidth]{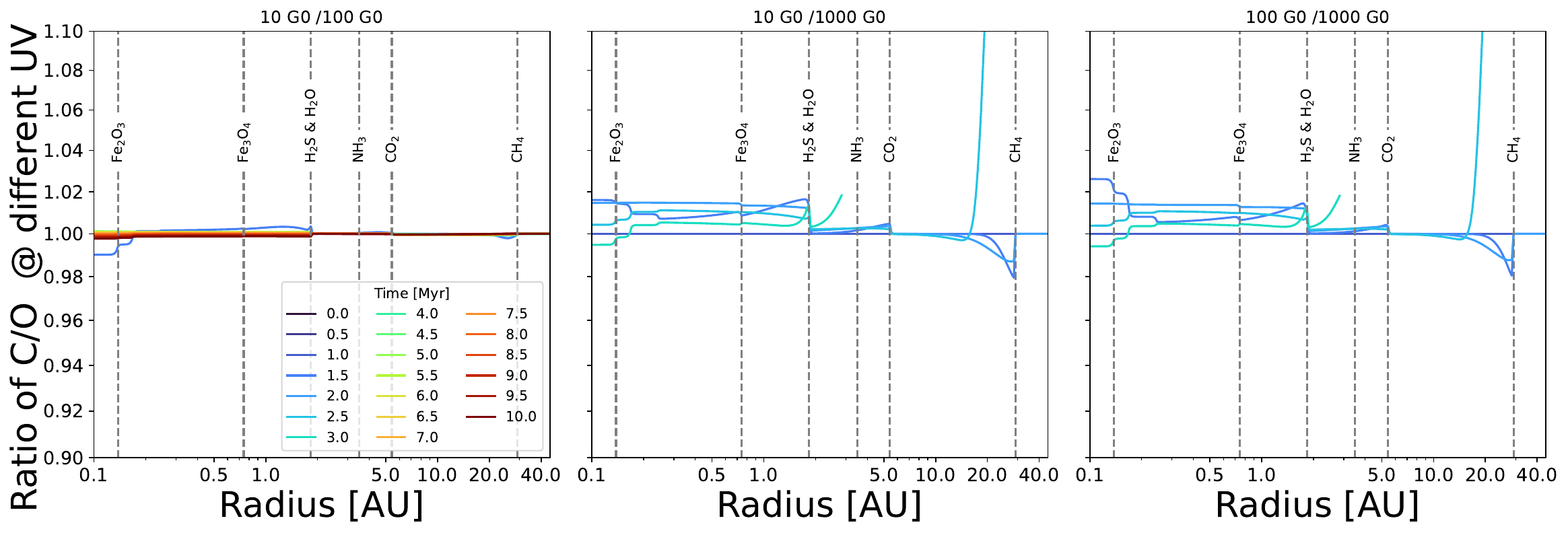}
   \caption{Comparison of C/O ratios across different UV environments as depicted in Fig~\ref{fig:CO_ratio_with_1000G0-nominal} but for the disc regions until 40~AU. The left, middle, and right panels illustrate the C/O ratio comparisons between UV environments of 10 $\mathbf{G}_0$ and 100 $\mathbf{G}_0$, 10 $\mathbf{G}_0$ and 1000 $\mathbf{G}_0$, and 100 $\mathbf{G}_0$ and 1000 $\mathbf{G}_0$, respectively.}
    \label{fig:UV_C/O_ratios}
    \end{figure*}

\subsection{Chemical evolution}\label{chemevolution}

The initial C/O ratio of the disc is set by our initial partitioning model, which describes how the volatiles are distributed in the disc (see appendix~\ref{app:parameters}). The C/O ratio changes at the different evaporation fronts and initial resembles the distribution as in \citet{2011ApJ...743L..16O} with a few extra chemical species. As time evolves, pebbles drift inwards and evaporate, changing the disc's C/O ratio (Fig.~\ref{fig:sig_gas}). Initially, the C/O ratio in the inner disc drops and becomes highly sub-stellar, due to the evaporation of water ice particles. Only after 6-7 Myr (due to the low viscosity) the C/O ratio becomes super-stellar again, due to the removal of oxygen rich material by accretion onto the star, and by inward flux of carbon rich materials on the longer viscous time-scales \citep{mahCloseinIceLines2023, 2024arXiv240606219M}. External photoevaporation will cut this process short. In the following detailed discussion, the focus is on the inner disc regions interior to the water ice evaporation front.

As in the pure viscous case, the C/O ratio initially in the disc drops due to the evaporation of inward drifting water ice (Fig.~\ref{cotevap0}). With time, as for the purely viscous case, the C/O ratio increases slowly, but it never reaches super-solar values, because the disc is dispersed by photoevaporation after $\sim$1.5 Myr or even earlier for low viscosities, preventing a significant change of the disc's C/O ratio before disc dissipation. For higher viscosities ($\alpha=10^{-3}$), the C/O ratio reaches roughly stellar values before the disc is dispersed. This is caused by the faster inward transportation of water vapour onto the star as well as by inward transport of carbon rich gas (see \citealt{mahCloseinIceLines2023, 2024arXiv240606219M}) by the larger viscosity.

If external photoevaporation becomes more efficient at later times (e.g. 1 or 3 Myr, see Fig.~\ref{cotevap1} and Fig.~\ref{cotevap3}), the disc's C/O ratio can evolve longer as in the purely viscous case, giving rise to larger C/O ratios in the inner disc. As long as the external photoevaporation is not active, the disc's evolution follows the purely viscous case (see Fig.~\ref{fig:sig_gas}). Generally speaking, a later start of external photoevaporation results in a C/O ratio evolution more similar to the purely viscous case. Only once external photoevaporation acts, the C/O ratio deviates from the purely viscous case. This implies for a relatively late starting time of external photoevaporation of 3 Myr that the inner disc regions can already become super-solar in C/O for high viscosity. However, this remains unachievable for the low viscosity discs, because of the longer evolution time.

Figure~\ref{fig:CH_OH_NH_with_ph} shows the variation in the ratios of carbon to hydrogen (C/H), oxygen to hydrogen (O/H), and nitrogen to hydrogen (N/H) over time in a protoplanetary disc that experiences external photoevaporation after 1 Myr of it's evolutionary time. The disc has a mass ($M_{\rm disc}$) of 0.1 solar masses, a radius ($R_{\rm disc}$) of 75 AU, and a viscosity parameter ($\alpha$) of $10^{-4}$. We show in Fig.~\ref{fig:CH_OH_NH_with_ph-nominal} the chemical evolution of a purely viscous discs for comparison.

The left panel of Fig.~\ref{fig:CH_OH_NH_with_ph} shows the changes in the C/H ratio over the disc life time. The C/H ratio in the disc increases at the various evaporation fronts that contain carbon bearing species. This is indicated by the sharp spikes in the distributions just interior to the evaporation fronts. The height of the spikes is determined by the overall abundances of the corresponding volatile ice. With time, the gas then slowly moves inwards, increasing the C/H ratio interior to the evaporation fronts. The inner disc's C/H ratio is then determined by the evaporation of inward drifting and evaporating CO$_2$ ice particles, reaching super-stellar levels before disc dissipation.

The middle panel of Fig.~\ref{fig:CH_OH_NH_with_ph} depicts the O/H ratio. Again a change in the disc's O/H ratio is determined by the evaporation of oxygen rich volatiles. As water ice is the most abundant oxygen carrying volatile, the jump in the disc's O/H ratio is largest at the water evaporation front. In fact the O/H ratio increases to about a factor of 20-30 interior to the water ice line. In combination with the C/H ratio that increased to around 2 times the stellar value, the disc's C/O ratio decreased to about $\left(\frac{1}{10}\right)^{\text{th}}$–~$\left(\frac{1}{15}\right)^{\text{th}}$ of its original value (see Fig.~\ref{fig:sig_gas}). Due to the low viscosity, the disc remains oxygen rich in the inner regions until the end of the disc's lifetime.

The right panel of Fig.~\ref{fig:CH_OH_NH_with_ph} shows the N/H ratio evolution as function of time. Here only N$_2$ and NH$_3$ carry nitrogen, resulting in only two evaporation fronts, where the N/H ratio increases. Interior the NH$_3$ evaporation front, the nitrogen content can increase to a few times stellar value, in line with the composition of Jupiter (see \citealt{schneiderHowDriftingEvaporating2021a}). Again, the inner disc remains enriched in nitrogen until the end of the disc's lifetime.

It is very clear that the chemical evolution of the disc first proceeds as in the nominal purely viscous case with rapid inward drifting pebbles that evaporate at ice lines and then enrich the disc with vapour. After the external photoevaporation starts to act, the disc's outer radius is cut in relatively short times, preventing further inward drift of pebbles and vapour. Consequently, the inner disc's composition remains relatively unchanged until the disc dissipates completely—explaining the similar inner disc composition observed in both isolated and externally photoevaporated discs (see Fig~\ref{fig:CH_OH_NH_with_ph} and Fig.~\ref{fig:CH_OH_NH_with_ph-nominal}) for comparison).

\subsection{Change of photoevaporative strength}\label{app:photo}

We show in Fig.~\ref{fig:CO_ratio_with_1000G0-nominal} the evolution of discs subject to external photoevaporation starting at 1~Myr. These scenarios involve a lower external field strength of \(10^{3} \mathbf{G}_0\),~\(10^{2} \mathbf{G}_0\), and \(10 \mathbf{G}_0\), which are factors of 10, 100 and 1000, respectively lower than those discussed in section~\ref{chemevolution}. The disc's viscosity is $\alpha=10^{-4}$, while the disc radius is 75 AU and the disc's mass is 0.1 solar masses, corresponding to our nominal disc case. We show in the figures the gas surface density evolution (top) and the evolution of the C/O ratio (bottom).

Due to the lower strength of external photoevaporation, the overall evolution is slower compared to the nominal cases with higher external photoevaporation rate. The lower external photoevaporation rates essentially prolong the disc's lifetime by about 1 Myr (Fig.~\ref{fig:CO_ratio_with_1000G0-nominal}--left plot), and beyond 7.5 Myr (Fig.~\ref{fig:CO_ratio_with_1000G0-nominal}--center and right plots) compared to the nominal case, leading to complete disc dissipation after 3.5~Myrs  (Fig.~\ref{fig:CO_ratio_with_1000G0-nominal}--left plot) or a disc survival beyond 10~Myrs (Fig.~\ref{fig:CO_ratio_with_1000G0-nominal}--center and right plots, similar to the purely viscous case. This confirms that at lower external photoevaporation, the lifetimes for discs are similar to those without external photoevaporation. 

The evolution of the disc's C/O ratio is similar to our previous simulations. The only difference is that the longer disc lifetime allows a larger increase of the C/O ratio in the inner disc, as there is more time available for the water vapour to be accreted onto the star and for carbon enhanced gas to move inwards. The overall values before external photoevaporations starts are in agreement with the purely viscous evolution (Fig.~\ref{fig:sig_gas}), as before. 

A low external photoevaporation rate essentially has no influence on the chemical composition of the inner disc—see Fig~\ref{fig:UV_C/O_ratios}. In Fig~\ref{fig:UV_C/O_ratios} we present the ratios of the disc C/O ratio for our 3 different external photoevaporation rates for disc regions until 40~AU. This plot illustrates that the C/O ratio in the inner disc is unaffected by the external photoevaporation rate, as long as the disc is not loosing mass in an effective manner. There is however difference in the outer disc chemistry for stronger UV environments—particularly around 20~AU and at disc times of around 2.5 Myr and later—where photoevaporation becomes efficient and starts to disrupt the disc (see Fig.~\ref{fig:UV_C/O_ratios}).

\section{Discussions}\label{sec:Compar}

As the protoplanetary disc evolves, small dust grains coalesce into pebbles and migrate inward. The growth of these grains correlates with the orbital period; hence, grains in the outer regions of the disc require more time to develop. Larger discs can therefore sustain a pebble flux into their central regions for extended periods \citep{bitschEnrichingInnerDiscs2023}. However, this process is moderated by external photoevaporation, which truncates the disc's outer parts, diminishing the influx of pebbles to the inner regions and potentially stunting planetary formation \citep{2023MNRAS.522.1939Q}. Our simulations confirm these previous results, as long as the external photoevaporation rate disperses the disc within the first Myr. If external photoevaporation acts later, planets could still form, as long as the embryos can start to accrete early (e.g. \citealt{2023A&A...679A..42S,2024arXiv240708533S} and appendix~\ref{app:flux}). However, these planets might then be hindered in their gas accretion phase and remain Neptune size cores. Planets that form in the inner regions of the disc are expected to have similar chemical compositions as in the purely viscous case, as the chemical evolution of the disc remains similar to the purely viscous disc evolution. The external photoevaporation thus only sets the lifetime of the disc and when the planet should stop growing, but not altering the planetary composition, justifying the approach taken in previous studies, where the disc is evaporated ad hoc (e.g. \citealt{bitschHowDriftingEvaporating2022}).

On the other hand, internal photoevaporation similarly plays a critical role by inducing gaps within protoplanetary discs. These gaps disrupt the homogeneous distribution of gas and dust, with gas diffusing into the gaps subsequently swept away by photoevaporative winds. Furthermore, substructures within the disc might reflect gap-opening processes\footnote{We remind the reader that not all substructures in discs correspond to gaps caused by growing planets—other physical processes may also be at play. For a detailed discussion, see \cite{2019MNRAS.488.3625N} or \cite{2023A&A...677A..82T}.} that act as dust traps, potentially explaining the redistribution of dust particles throughout the disc \citep{2024A&A...681A..84G}. These processes effectively block the inward drift of gas and pebbles, complicating the attainment of solar or super-solar carbon-to-oxygen (C/O) ratios in the inner disc regions \citep{2024arXiv240209342L}. Our results align with this perspective, pinpointing how both external and internal photoevaporation uniformly influence the chemical composition of the inner disc, despite their different underlying mechanisms.

Recent observations by \cite{2024arXiv240500615D} identified CO isotopologues in various protoplanetary discs. These findings indicate that the discs, featuring moderately irradiated environments, showed no substantial differences in CO abundances between isolated and irradiated settings. The study suggests that the observed chemical similarities are likely due to variations in stellar mass rather than the influence of external far-ultraviolet (FUV) radiation. This indicates that the discs are located at distances where the massive Trapezium stars have minimal impact on their chemical compositions.

On the other hand, \citet{2023ApJ...958L..30R} studied a disc in the NGC 6357 region, where extreme UV environments from close-by numerous massive OB stars are present. Studying a young disc (age < 1 Myr) in this region, \citet{2023ApJ...958L..30R} find no significant difference in the chemical composition of the inner disc regions compared to more isolated discs that also show a significant water excess (e.g. \citealt{perottiWaterTerrestrialPlanetforming2023, grantMINDSDetection132023, banzattiJWSTRevealsExcess2023}). Our simulations clearly show that external photoevaporation, even though it dispersed away the very outer disc regions, does not influence significantly the inner disc's composition, as the pebbles already have a few 100~kyrs—1~Myr to drift inward and enrich the disc with vapour, maintaining a low C/O ratio, before external photoevaporation acts. Furthermore, our model implies that a low viscosity is needed, as larger viscosities (e.g.$\alpha=10^{-3}$) would result in an already close to solar C/O ratio, before external photoevaporation becomes efficient.

Given these insights, our findings suggest important implications for future observations aimed at observing chemical abundances in highly irradiated (and non-irradiated) regions of inner discs. Our model predicts that the C/O ratio in these inner discs should be low (as long as the viscosity is low) as long as the discs are young. Consequently, we'd expect larger C/O ratios for older discs. On the other hand, the C/O ratio of the inner disc region can be influenced by outer pressure perturbations \citep{2024arXiv240606219M}, showing the need to get a full picture of the disc's structure and composition to constrain it's evolution.

\section{Conclusions}\label{sec:summary}

In this study, we have conducted 1D semi-analytical simulations of protoplanetary discs irradiated by external far-ultraviolet (FUV) radiation, focusing on systems hosted by solar-like stars. Our model incorporates both pebble drift and the effects of pebble evaporation. We compare the results of our simulations with external photoevaporation to a purely viscous case.

Our findings demonstrate that external photoevaporation exerts a profound influence on the evolution of protoplanetary discs, in line with previous studies (e.g. \citealt{2023MNRAS.522.1939Q}). Specifically, external photoevaporation truncates the outer regions of the disc, effectively halting the influx of pebbles and thereby impeding the formation of planetary cores within the inner disc, as long as external photoevaporation starts to act early. If external photoevaporation becomes efficient only after 1-2 Myr, the main amount of pebbles already drifted inwards (see appendix~\ref{app:flux}), giving rise to efficient planet formation (e.g. \citealt{2023A&A...679A..42S}). On the other hand, low mass discs are dissipated by external photoevaporation as soon as it starts, destroying all hopes to form planets. However, these small mass discs would have had trouble to form giant planets in the first place (e.g. \citealt{2023A&A...679A..42S}).

The inner disc composition is driven by inward drifting and evaporating pebbles. As the pebbles grow quickly and move inwards on short time-scales, they can enrich the inner disc already very early (see also \citealt{2024arXiv240606219M}), resulting in sub-solar C/O ratios with super-solar water contents. Over time, this sub-solar C/O ratio increases due to the accretion of water vapour onto the star and inward moving carbon rich gas. Once external photoevaporation starts to act, the inner disc chemistry continues to evolve as in the viscous case, because the enriched gas in the inner few AU is not influenced by the removal of material at a few 10 AUs. As external photoevaporation continues the whole disc is eventually dispersed away. 

In our model, the discs remain sub-solar in their C/O ratio for a few Myr, similar to the pure viscous case, before external photoevaporation removes the disc. In this case our model predicts that inner discs should have a low C/O ratio and a high water content, even when subject to external photoevaporation, in line with observations \citep{2023ApJ...958L..30R}. 

On the other hand, higher viscosities could allow larger C/O ratios in the inner disc region, because the water vapour is accreted onto the star on shorter time-scales and carbon rich material can make its way inwards faster. Nevertheless, the chemical evolution is again very similar to a purely viscous disc, even after external photoevaporation starts to act. This implies that the influence of external photoevaporation on the composition of planets residing in the inner discs might be minimal. Furthermore, we predict that there should be no difference in the chemical compositions of discs that are subject to external photoevaporation compared to discs that are in low radiation environment (aka isolated environments), where external photoevaporation plays only a minor role, testable with future JWST observations (e.g. \citealt{2023ApJ...958L..30R}).

\section*{Acknowledgements}
We thank Maria-Claudia Ramirez-Tannus for the helpful discussions. We also thank the anonymous referee whose comments substantially improved this paper.

\section*{Funding Statement}
N.N. acknowledges the financial support of Max-Planck Africa Mobility Grant that helped to set up this research collaboration.

\bibliographystyle{aa} 
\bibliography{Paper_1} 

\begin{thebibliography}{84}
\expandafter\ifx\csname natexlab\endcsname\relax\def\natexlab#1{#1}\fi

\bibitem[{Armitage(2013)}]{armitageAstrophysicsPlanetFormation2013}
Armitage, P.~J. 2013, Astrophysics of {{Planet Formation}} ({Cambridge
  University Press})

\bibitem[{Asplund {et~al.}(2009)Asplund, Grevesse, Sauval, \&
  Scott}]{asplundChemicalCompositionSun2009}
Asplund, M., Grevesse, N., Sauval, A.~J., \& Scott, P. 2009, Annual Review of
  Astronomy and Astrophysics, 47, 481

\bibitem[{Ataiee {et~al.}(2018)Ataiee, Baruteau, Alibert, \&
  Benz}]{ataieeHowMuchDoes2018}
Ataiee, S., Baruteau, C., Alibert, Y., \& Benz, W. 2018, Astronomy \&
  Astrophysics, 615, A110

\bibitem[{Banzatti {et~al.}(2023)Banzatti, Pontoppidan, Carr, Jellison,
  Pascucci, Najita, {Munoz-Romero}, Oberg, Kalyaan, Pinilla, Krijt, Long,
  Lambrechts, Rosotti, Herczeg, Salyk, Zhang, Bergin, Ballering, Meyer,
  Bruderer, \& {collaboration}}]{banzattiJWSTRevealsExcess2023}
Banzatti, A., Pontoppidan, K.~M., Carr, J., {et~al.} 2023

\bibitem[{{Bergez-Casalou} {et~al.}(2020){Bergez-Casalou}, {Bitsch}, {Pierens},
  {Crida}, \& {Raymond}}]{2020A&A...643A.133B}
{Bergez-Casalou}, C., {Bitsch}, B., {Pierens}, A., {Crida}, A., \& {Raymond},
  S.~N. 2020, \aap, 643, A133

\bibitem[{Birnstiel {et~al.}(2012)Birnstiel, Klahr, \&
  Ercolano}]{birnstielSimpleModelEvolution2012}
Birnstiel, T., Klahr, H., \& Ercolano, B. 2012, Astronomy \& Astrophysics, 539,
  A148

\bibitem[{Bitsch \& Battistini(2020)}]{bitschInfluenceSubSupersolar2020}
Bitsch, B. \& Battistini, C. 2020, Astronomy \& Astrophysics, 633, A10

\bibitem[{{Bitsch} {et~al.}(2015){Bitsch}, {Lambrechts}, \&
  {Johansen}}]{2015A&A...582A.112B}
{Bitsch}, B., {Lambrechts}, M., \& {Johansen}, A. 2015, \aap, 582, A112

\bibitem[{Bitsch \& Mah(2023)}]{bitschEnrichingInnerDiscs2023}
Bitsch, B. \& Mah, J. 2023

\bibitem[{Bitsch {et~al.}(2018)Bitsch, Morbidelli, Johansen, Lega, Lambrechts,
  \& Crida}]{bitschPebbleisolationMassScaling2018}
Bitsch, B., Morbidelli, A., Johansen, A., {et~al.} 2018, Astronomy \&
  Astrophysics, 612, A30

\bibitem[{Bitsch {et~al.}(2021)Bitsch, Raymond, Buchhave, {Bello-Arufe},
  Rathcke, \& Schneider}]{bitschDryWaterWorld2021}
Bitsch, B., Raymond, S.~N., Buchhave, L.~A., {et~al.} 2021, Astronomy \&
  Astrophysics, 649, L5

\bibitem[{Bitsch {et~al.}(2022)Bitsch, Schneider, \&
  Kreidberg}]{bitschHowDriftingEvaporating2022}
Bitsch, B., Schneider, A.~D., \& Kreidberg, L. 2022, Astronomy \& Astrophysics,
  665, A138

\bibitem[{{Bitsch} {et~al.}(2022){Bitsch}, {Schneider}, \&
  {Kreidberg}}]{2022A&A...665A.138B}
{Bitsch}, B., {Schneider}, A.~D., \& {Kreidberg}, L. 2022, \aap, 665, A138

\bibitem[{{Blum} \& {Wurm}(2008)}]{2008ARA&A..46...21B}
{Blum}, J. \& {Wurm}, G. 2008, \araa, 46, 21

\bibitem[{{Booth} {et~al.}(2017){Booth}, {Clarke}, {Madhusudhan}, \&
  {Ilee}}]{2017MNRAS.469.3994B}
{Booth}, R.~A., {Clarke}, C.~J., {Madhusudhan}, N., \& {Ilee}, J.~D. 2017,
  \mnras, 469, 3994

\bibitem[{Booth \& Ilee(2019)}]{boothPlanetformingMaterialProtoplanetary2019}
Booth, R.~A. \& Ilee, J.~D. 2019, Monthly Notices of the Royal Astronomical
  Society, 487, 3998

\bibitem[{Brauer {et~al.}(2008)Brauer, Dullemond, \&
  Henning}]{brauerCoagulationFragmentationRadial2008}
Brauer, F., Dullemond, C.~P., \& Henning, {\relax Th}. 2008, Astronomy \&
  Astrophysics, 480, 859

\bibitem[{{Chiang} \& {Goldreich}(1997)}]{1997ApJ...490..368C}
{Chiang}, E.~I. \& {Goldreich}, P. 1997, \apj, 490, 368

\bibitem[{{D'Angelo} \& {Lubow}(2008)}]{2008ApJ...685..560D}
{D'Angelo}, G. \& {Lubow}, S.~H. 2008, \apj, 685, 560

\bibitem[{{D{\'\i}az-Berr{\'\i}os} {et~al.}(2024){D{\'\i}az-Berr{\'\i}os},
  {Guzm{\'a}n}, {Walsh}, {{\"O}berg}, {Cleeves}, {de la Villarmois}, \&
  {Carpenter}}]{2024arXiv240500615D}
{D{\'\i}az-Berr{\'\i}os}, J.~K., {Guzm{\'a}n}, V.~V., {Walsh}, C., {et~al.}
  2024, arXiv e-prints, arXiv:2405.00615

\bibitem[{{Dr{\k{a}}{\.z}kowska} {et~al.}(2019){Dr{\k{a}}{\.z}kowska}, {Li},
  {Birnstiel}, {Stammler}, \& {Li}}]{2019ApJ...885...91D}
{Dr{\k{a}}{\.z}kowska}, J., {Li}, S., {Birnstiel}, T., {Stammler}, S.~M., \&
  {Li}, H. 2019, \apj, 885, 91

\bibitem[{{Ercolano} {et~al.}(2008){Ercolano}, {Drake}, {Raymond}, \&
  {Clarke}}]{2008ApJ...688..398E}
{Ercolano}, B., {Drake}, J.~J., {Raymond}, J.~C., \& {Clarke}, C.~C. 2008,
  \apj, 688, 398

\bibitem[{{Estrada} \& {Cuzzi}(2022)}]{2022ApJ...936...40E}
{Estrada}, P.~R. \& {Cuzzi}, J.~N. 2022, \apj, 936, 40

\bibitem[{{Estrada} {et~al.}(2022){Estrada}, {Cuzzi}, \&
  {Umurhan}}]{2022ApJ...936...42E}
{Estrada}, P.~R., {Cuzzi}, J.~N., \& {Umurhan}, O.~M. 2022, \apj, 936, 42

\bibitem[{{Fatuzzo} \& {Adams}(2008)}]{2008ApJ...675.1361F}
{Fatuzzo}, M. \& {Adams}, F.~C. 2008, \apj, 675, 1361

\bibitem[{{G{\'a}rate} {et~al.}(2024){G{\'a}rate}, {Pinilla}, {Haworth}, \&
  {Facchini}}]{2024A&A...681A..84G}
{G{\'a}rate}, M., {Pinilla}, P., {Haworth}, T.~J., \& {Facchini}, S. 2024,
  \aap, 681, A84

\bibitem[{Gasman {et~al.}(2023)Gasman, Van~Dishoeck, Grant, Temmink, Tabone,
  Henning, Kamp, G{\"u}del, Lagage, Perotti, Christiaens, Samland, Arabhavi,
  Argyriou, Abergel, Absil, Barrado, Boccaletti, Bouwman, Caratti O~Garatti,
  Geers, Glauser, Guadarrama, Jang, Kanwar, Lahuis, {Morales-Calder{\'o}n},
  Mueller, Nehm{\'e}, Olofsson, Pantin, Pawellek, Ray, {Rodgers-Lee},
  Scheithauer, Schreiber, Schwarz, Vandenbussche, Vlasblom, Waters, Wright,
  Colina, Greve, \& {\"O}stlin}]{gasmanMINDSAbundantWater2023}
Gasman, D., Van~Dishoeck, E.~F., Grant, S.~L., {et~al.} 2023, Astronomy \&
  Astrophysics, 679, A117

\bibitem[{{Ginski} {et~al.}(2024){Ginski}, {Garufi}, {Benisty}, {Tazaki},
  {Dominik}, {Ribas}, {Engler}, {Birnstiel}, {Chauvin}, {Columba}, {Facchini},
  {Goncharov}, {Hagelberg}, {Henning}, {Hogerheijde}, {van Holstein}, {Huang},
  {Muto}, {Pinilla}, {Kanagawa}, {Kim}, {Kurtovic}, {Langlois}, {Manara},
  {Milli}, {Momose}, {Orihara}, {Pawellek}, {Pinte}, {Rab}, {Schmidt}, {Snik},
  {Wahhaj}, {Williams}, \& {Zurlo}}]{2024arXiv240302149G}
{Ginski}, C., {Garufi}, A., {Benisty}, M., {et~al.} 2024, arXiv e-prints,
  arXiv:2403.02149

\bibitem[{Grant {et~al.}(2023)Grant, Van~Dishoeck, Tabone, Gasman, Henning,
  Kamp, G{\"u}del, Lagage, Bettoni, Perotti, Christiaens, Samland, Arabhavi,
  Argyriou, Abergel, Absil, Barrado, Boccaletti, Bouwman, O~Garatti, Geers,
  Glauser, Guadarrama, Jang, Kanwar, Lahuis, {Morales-Calder{\'o}n}, Mueller,
  Nehm{\'e}, Olofsson, Pantin, Pawellek, Ray, {Rodgers-Lee}, Scheithauer,
  Schreiber, Schwarz, Temmink, Vandenbussche, Vlasblom, Waters, Wright, Colina,
  Greve, Justannont, \& {\"O}stlin}]{grantMINDSDetection132023}
Grant, S.~L., Van~Dishoeck, E.~F., Tabone, B., {et~al.} 2023, The Astrophysical
  Journal Letters, 947, L6

\bibitem[{Gundlach \&
  Blum(2014)}]{gundlachSTICKINESSMICROMETERSIZEDWATERICE2014}
Gundlach, B. \& Blum, J. 2014, The Astrophysical Journal, 798, 34

\bibitem[{{G{\"u}ttler} {et~al.}(2010){G{\"u}ttler}, {Blum}, {Zsom}, {Ormel},
  \& {Dullemond}}]{2010A&A...513A..56G}
{G{\"u}ttler}, C., {Blum}, J., {Zsom}, A., {Ormel}, C.~W., \& {Dullemond},
  C.~P. 2010, \aap, 513, A56

\bibitem[{{Haworth} \& {Clarke}(2019)}]{2019MNRAS.485.3895H}
{Haworth}, T.~J. \& {Clarke}, C.~J. 2019, \mnras, 485, 3895

\bibitem[{{Haworth} {et~al.}(2023){Haworth}, {Coleman}, {Qiao}, {Sellek}, \&
  {Askari}}]{2023MNRAS.526.4315H}
{Haworth}, T.~J., {Coleman}, G. A.~L., {Qiao}, L., {Sellek}, A.~D., \&
  {Askari}, K. 2023, \mnras, 526, 4315

\bibitem[{{Henney} \& {Arthur}(1998)}]{1998AJ....116..322H}
{Henney}, W.~J. \& {Arthur}, S.~J. 1998, \aj, 116, 322

\bibitem[{{Henney} \& {O'Dell}(1999)}]{1999AJ....118.2350H}
{Henney}, W.~J. \& {O'Dell}, C.~R. 1999, \aj, 118, 2350

\bibitem[{{Johansen} {et~al.}(2014){Johansen}, {Blum}, {Tanaka}, {Ormel},
  {Bizzarro}, \& {Rickman}}]{2014prpl.conf..547J}
{Johansen}, A., {Blum}, J., {Tanaka}, H., {et~al.} 2014, in Protostars and
  Planets VI, ed. H.~{Beuther}, R.~S. {Klessen}, C.~P. {Dullemond}, \&
  T.~{Henning}, 547--570

\bibitem[{Kalyaan {et~al.}(2023)Kalyaan, Pinilla, Krijt, Banzatti, Rosotti,
  Mulders, Lambrechts, Long, \& Herczeg}]{kalyaanEffectDustEvolution2023}
Kalyaan, A., Pinilla, P., Krijt, S., {et~al.} 2023, The Astrophysical Journal,
  954, 66

\bibitem[{Kalyaan {et~al.}(2021)Kalyaan, Pinilla, Krijt, Mulders, \&
  Banzatti}]{kalyaanLinkingOuterDisk2021}
Kalyaan, A., Pinilla, P., Krijt, S., Mulders, G.~D., \& Banzatti, A. 2021, The
  Astrophysical Journal, 921, 84

\bibitem[{{Lada} \& {Lada}(2003)}]{2003ARA&A..41...57L}
{Lada}, C.~J. \& {Lada}, E.~A. 2003, \araa, 41, 57

\bibitem[{{Lambrechts} \& {Johansen}(2014)}]{2014A&A...572A.107L}
{Lambrechts}, M. \& {Johansen}, A. 2014, \aap, 572, A107

\bibitem[{Lambrechts {et~al.}(2014)Lambrechts, Johansen, \&
  Morbidelli}]{lambrechtsSeparatingGasgiantIcegiant2014}
Lambrechts, M., Johansen, A., \& Morbidelli, A. 2014, Astronomy \&
  Astrophysics, 572, A35

\bibitem[{{Lienert} {et~al.}(2024){Lienert}, {Bitsch}, \&
  {Henning}}]{2024arXiv240209342L}
{Lienert}, J.~L., {Bitsch}, B., \& {Henning}, T. 2024, arXiv e-prints,
  arXiv:2402.09342

\bibitem[{Lodders(2003)}]{loddersSolarSystemAbundances2003}
Lodders, K. 2003, The Astrophysical Journal, 591, 1220

\bibitem[{{Lynden-Bell} \&
  Pringle(1974)}]{lynden-bellEvolutionViscousDiscs1974a}
{Lynden-Bell}, D. \& Pringle, J.~E. 1974, Monthly Notices of the Royal
  Astronomical Society, 168, 603

\bibitem[{Madhusudhan {et~al.}(2014)Madhusudhan, Knutson, Fortney, \&
  Barman}]{madhusudhanExoplanetaryAtmospheres2014}
Madhusudhan, N., Knutson, H., Fortney, J.~J., \& Barman, T. 2014, in Protostars
  and {{Planets VI}} ({University of Arizona Press})

\bibitem[{Mah {et~al.}(2023)Mah, Bitsch, Pascucci, \&
  Henning}]{mahCloseinIceLines2023}
Mah, J., Bitsch, B., Pascucci, I., \& Henning, T. 2023, Astronomy \&
  Astrophysics, 677, L7

\bibitem[{{Mah} {et~al.}(2024){Mah}, {Savvidou}, \&
  {Bitsch}}]{2024arXiv240606219M}
{Mah}, J., {Savvidou}, S., \& {Bitsch}, B. 2024, arXiv e-prints,
  arXiv:2406.06219

\bibitem[{Molli{\`e}re {et~al.}(2022)Molli{\`e}re, Molyarova, Bitsch, Henning,
  Schneider, Kreidberg, Eistrup, Burn, Nasedkin, Semenov, Mordasini, Schlecker,
  Schwarz, Lacour, Nowak, \&
  Schulik}]{molliereInterpretingAtmosphericComposition2022}
Molli{\`e}re, P., Molyarova, T., Bitsch, B., {et~al.} 2022, The Astrophysical
  Journal, 934, 74

\bibitem[{Mousis {et~al.}(2022)Mousis, Atkinson, \& {Ice Giants
  team}}]{mousisSituExplorationAtmospheres2022}
Mousis, O., Atkinson, D.~H., \& {Ice Giants team}. 2022, In {{Situ
  Exploration}} of the Atmospheres of the {{Ice Giants}}, Other, {display}

\bibitem[{Mousis {et~al.}(2009)Mousis, Marboeuf, Lunine, Alibert, Fletcher,
  Orton, Pauzat, \& Ellinger}]{mousisDETERMINATIONMINIMUMMASSES2009}
Mousis, O., Marboeuf, U., Lunine, J.~I., {et~al.} 2009, The Astrophysical
  Journal, 696, 1348

\bibitem[{{M{\"u}ller} {et~al.}(2021){M{\"u}ller}, {Savvidou}, \&
  {Bitsch}}]{2021A&A...650A.185M}
{M{\"u}ller}, J., {Savvidou}, S., \& {Bitsch}, B. 2021, \aap, 650, A185

\bibitem[{Musiolik \& Wurm(2019)}]{musiolikContactsWaterIce2019}
Musiolik, G. \& Wurm, G. 2019, The Astrophysical Journal, 873, 58

\bibitem[{{Ndugu} {et~al.}(2022){Ndugu}, {Abedigamba}, \&
  {Andama}}]{2022MNRAS.512..861N}
{Ndugu}, N., {Abedigamba}, O.~P., \& {Andama}, G. 2022, \mnras, 512, 861

\bibitem[{Ndugu {et~al.}(2018)Ndugu, Bitsch, \&
  Jurua}]{nduguPlanetPopulationSynthesis2018}
Ndugu, N., Bitsch, B., \& Jurua, E. 2018, Monthly Notices of the Royal
  Astronomical Society, 474, 886

\bibitem[{{Ndugu} {et~al.}(2019){Ndugu}, {Bitsch}, \&
  {Jurua}}]{2019MNRAS.488.3625N}
{Ndugu}, N., {Bitsch}, B., \& {Jurua}, E. 2019, \mnras, 488, 3625

\bibitem[{{{\"O}berg} {et~al.}(2011){{\"O}berg}, {Murray-Clay}, \&
  {Bergin}}]{2011ApJ...743L..16O}
{{\"O}berg}, K.~I., {Murray-Clay}, R., \& {Bergin}, E.~A. 2011, \apjl, 743, L16

\bibitem[{Owen {et~al.}(2012)Owen, Clarke, \&
  Ercolano}]{owenTheoryDiscPhotoevaporation2012a}
Owen, J.~E., Clarke, C.~J., \& Ercolano, B. 2012, Monthly Notices of the Royal
  Astronomical Society, 422, 1880

\bibitem[{Paardekooper \& Mellema(2006)}]{paardekooperDustFlowGas2006}
Paardekooper, S.-J. \& Mellema, G. 2006, Astronomy \& Astrophysics, 453, 1129

\bibitem[{Pascucci {et~al.}(2022)Pascucci, Cabrit, Edwards, Gorti, Gressel, \&
  Suzuki}]{pascucciRoleDiskWinds2022}
Pascucci, I., Cabrit, S., Edwards, S., {et~al.} 2022

\bibitem[{Perotti {et~al.}(2023)Perotti, Christiaens, Henning, Tabone, Waters,
  Kamp, Olofsson, Grant, Gasman, Bouwman, Samland, Franceschi, Van~Dishoeck,
  Schwarz, G{\"u}del, Lagage, Ray, Vandenbussche, Abergel, Absil, Arabhavi,
  Argyriou, Barrado, Boccaletti, Caratti O~Garatti, Geers, Glauser, Justannont,
  Lahuis, Mueller, Nehm{\'e}, Pantin, Scheithauer, Waelkens, Guadarrama, Jang,
  Kanwar, {Morales-Calder{\'o}n}, Pawellek, {Rodgers-Lee}, Schreiber, Colina,
  Greve, {\"O}stlin, \& Wright}]{perottiWaterTerrestrialPlanetforming2023}
Perotti, G., Christiaens, V., Henning, {\relax Th}., {et~al.} 2023, Nature,
  620, 516

\bibitem[{Picogna {et~al.}(2019)Picogna, Ercolano, Owen, \&
  Weber}]{picognaDispersalProtoplanetaryDiscs2019a}
Picogna, G., Ercolano, B., Owen, J.~E., \& Weber, M.~L. 2019, Monthly Notices
  of the Royal Astronomical Society, 487, 691

\bibitem[{{Picogna} {et~al.}(2019){Picogna}, {Ercolano}, {Owen}, \&
  {Weber}}]{2019MNRAS.487..691P}
{Picogna}, G., {Ercolano}, B., {Owen}, J.~E., \& {Weber}, M.~L. 2019, \mnras,
  487, 691

\bibitem[{Pinilla {et~al.}(2012)Pinilla, Birnstiel, Ricci, Dullemond, Uribe,
  Testi, \& Natta}]{pinillaTrappingDustParticles2012}
Pinilla, P., Birnstiel, T., Ricci, L., {et~al.} 2012, Astronomy \&
  Astrophysics, 538, A114

\bibitem[{Pinilla {et~al.}(2017)Pinilla, Pohl, Stammler, \&
  Birnstiel}]{Pinilla_2017}
Pinilla, P., Pohl, A., Stammler, S.~M., \& Birnstiel, T. 2017, The
  Astrophysical Journal, 845, 68

\bibitem[{{Piso} {et~al.}(2015){Piso}, {{\"O}berg}, {Birnstiel}, \&
  {Murray-Clay}}]{2015ApJ...815..109P}
{Piso}, A.-M.~A., {{\"O}berg}, K.~I., {Birnstiel}, T., \& {Murray-Clay}, R.~A.
  2015, \apj, 815, 109

\bibitem[{Pringle(1981)}]{pringleAccretionDiscsAstrophysics1981}
Pringle, J.~E. 1981, Annual Review of Astronomy and Astrophysics, 19, 137

\bibitem[{{Qiao} {et~al.}(2023){Qiao}, {Coleman}, \&
  {Haworth}}]{2023MNRAS.522.1939Q}
{Qiao}, L., {Coleman}, G. A.~L., \& {Haworth}, T.~J. 2023, \mnras, 522, 1939

\bibitem[{{Qiao} {et~al.}(2022){Qiao}, {Haworth}, {Sellek}, \&
  {Ali}}]{2022MNRAS.512.3788Q}
{Qiao}, L., {Haworth}, T.~J., {Sellek}, A.~D., \& {Ali}, A.~A. 2022, \mnras,
  512, 3788

\bibitem[{{Ram{\'\i}rez-Tannus} {et~al.}(2023){Ram{\'\i}rez-Tannus}, {Bik},
  {Cuijpers}, {Waters}, {G{\"o}ppl}, {Henning}, {Kamp}, {Preibisch}, {Getman},
  {Chaparro}, {Cuartas-Restrepo}, {de Koter}, {Feigelson}, {Grant}, {Haworth},
  {Hern{\'a}ndez}, {Kuhn}, {Perotti}, {Povich}, {Reiter}, {Roccatagliata},
  {Sabbi}, {Tabone}, {Winter}, {McLeod}, {van Boekel}, \& {van
  Terwisga}}]{2023ApJ...958L..30R}
{Ram{\'\i}rez-Tannus}, M.~C., {Bik}, A., {Cuijpers}, L., {et~al.} 2023, \apjl,
  958, L30

\bibitem[{{Ros} \& {Johansen}(2013)}]{2013A&A...552A.137R}
{Ros}, K. \& {Johansen}, A. 2013, \aap, 552, A137

\bibitem[{{Safronov}(1972)}]{1972epcf.book.....S}
{Safronov}, V.~S. 1972, {Evolution of the protoplanetary cloud and formation of
  the earth and planets.}

\bibitem[{{Savvidou} \& {Bitsch}(2023)}]{2023A&A...679A..42S}
{Savvidou}, S. \& {Bitsch}, B. 2023, \aap, 679, A42

\bibitem[{{Savvidou} \& {Bitsch}(2024)}]{2024arXiv240708533S}
{Savvidou}, S. \& {Bitsch}, B. 2024, arXiv e-prints, arXiv:2407.08533

\bibitem[{Schneider \&
  Bitsch(2021{\natexlab{a}})}]{schneiderHowDriftingEvaporating2021}
Schneider, A.~D. \& Bitsch, B. 2021{\natexlab{a}}, Astronomy \& Astrophysics,
  654, A71

\bibitem[{Schneider \&
  Bitsch(2021{\natexlab{b}})}]{schneiderHowDriftingEvaporating2021a}
Schneider, A.~D. \& Bitsch, B. 2021{\natexlab{b}}, Astronomy \& Astrophysics,
  654, A72

\bibitem[{{Schneider} \& {Bitsch}(2023)}]{2024arXiv240115686S}
{Schneider}, A.~D. \& {Bitsch}, B. 2023, arXiv e-prints, arXiv:2401.15686

\bibitem[{{Sellek} {et~al.}(2020){Sellek}, {Booth}, \&
  {Clarke}}]{2020MNRAS.492.1279S}
{Sellek}, A.~D., {Booth}, R.~A., \& {Clarke}, C.~J. 2020, \mnras, 492, 1279

\bibitem[{Shakura \& Sunyaev(1973)}]{shakuraBlackHolesBinary1973}
Shakura, N.~I. \& Sunyaev, R.~A. 1973, Astronomy and Astrophysics, 337

\bibitem[{{Stammler} {et~al.}(2023){Stammler}, {Lichtenberg},
  {Dr{\k{a}}{\.z}kowska}, \& {Birnstiel}}]{2023A&A...670L...5S}
{Stammler}, S.~M., {Lichtenberg}, T., {Dr{\k{a}}{\.z}kowska}, J., \&
  {Birnstiel}, T. 2023, \aap, 670, L5

\bibitem[{{Tzouvanou} {et~al.}(2023){Tzouvanou}, {Bitsch}, \&
  {Pichierri}}]{2023A&A...677A..82T}
{Tzouvanou}, A., {Bitsch}, B., \& {Pichierri}, G. 2023, \aap, 677, A82

\bibitem[{Weber {et~al.}(2018)Weber, {Ben{\'i}tez-Llambay}, Gressel, Krapp, \&
  Pessah}]{weberCharacterizingVariableDust2018}
Weber, P., {Ben{\'i}tez-Llambay}, P., Gressel, O., Krapp, L., \& Pessah, M.~E.
  2018, The Astrophysical Journal, 854, 153

\bibitem[{Weidenschilling(1977)}]{weidenschillingAerodynamicsSolidBodies1977}
Weidenschilling, S.~J. 1977, Monthly Notices of the Royal Astronomical Society,
  180, 57

\bibitem[{{Weidenschilling} \& {Cuzzi}(1993)}]{1993prpl.conf.1031W}
{Weidenschilling}, S.~J. \& {Cuzzi}, J.~N. 1993, in Protostars and Planets III,
  ed. E.~H. {Levy} \& J.~I. {Lunine}, 1031

\bibitem[{{Winter} {et~al.}(2020){Winter}, {Kruijssen}, {Longmore}, \&
  {Chevance}}]{2020Natur.586..528W}
{Winter}, A.~J., {Kruijssen}, J.~M.~D., {Longmore}, S.~N., \& {Chevance}, M.
  2020, \nat, 586, 528

\end{thebibliography}

\appendix

\section{Additional Information} \label{app:parameters}
   \begin{figure*}[h]
   \centering
   \includegraphics[width=\textwidth]{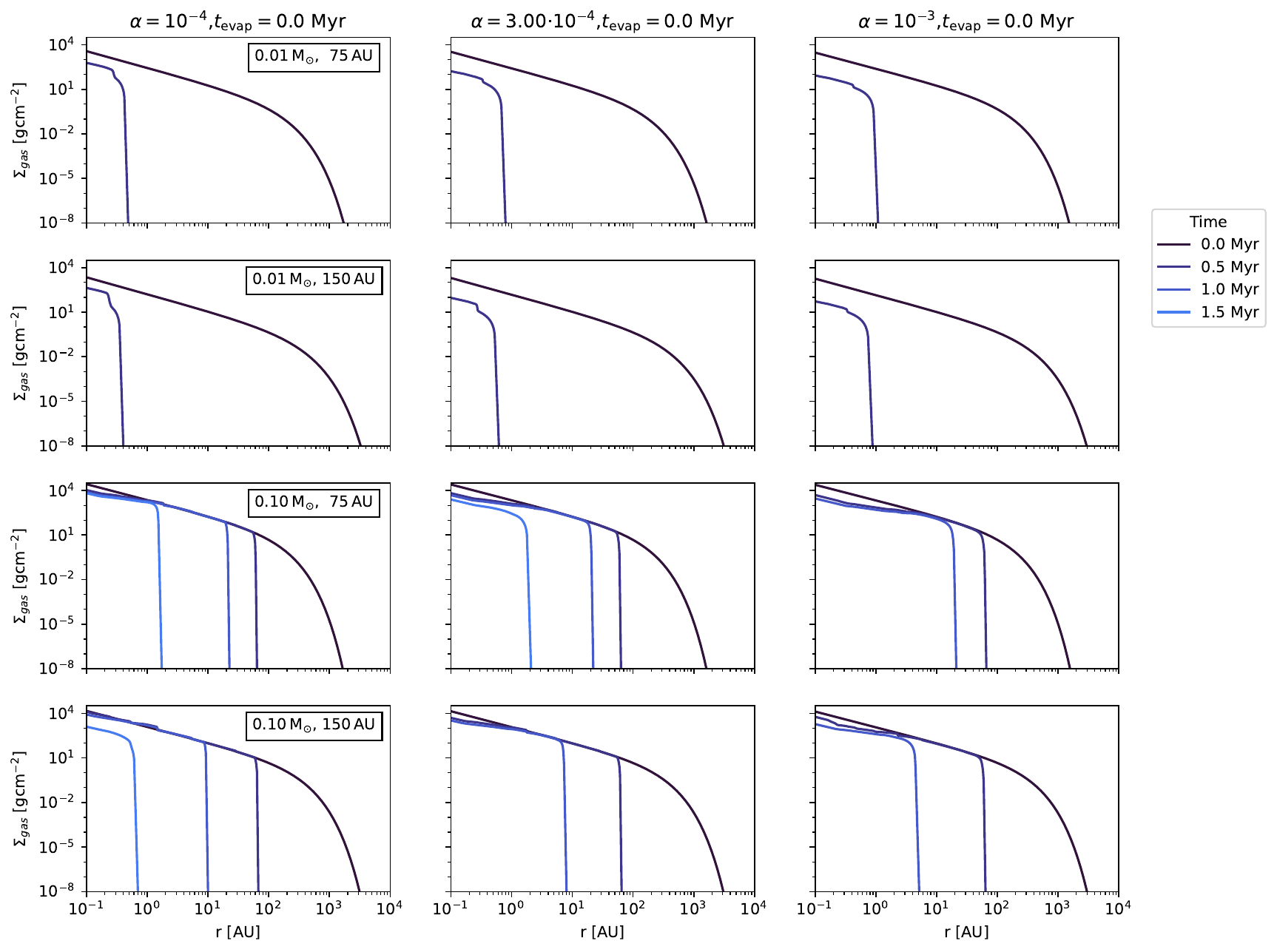}
   \caption{Gas surface density for a viscous disc with external photoevaporation from neighbouring stars. The gas surface density is shown as a function of disc radius and time. A variety of parameter configurations is used for this study: Along the x-axis of this figure, $\alpha$ is varied from $\alpha=10^{-4}$ to $\alpha=10^{-2}$, whereas on the y-axis, different sets of disc masses and radii are used, as indicated in the panels. The colour coding gives the time evolution. The discs with 0.01 solar masses are dispersed by external photoevaporation within the first 0.5 Myr and thus show now evolution in the plot.}
              \label{tevap0}%
    \end{figure*}

\begin{figure*}[h]
   \centering
   \includegraphics[width=\textwidth]{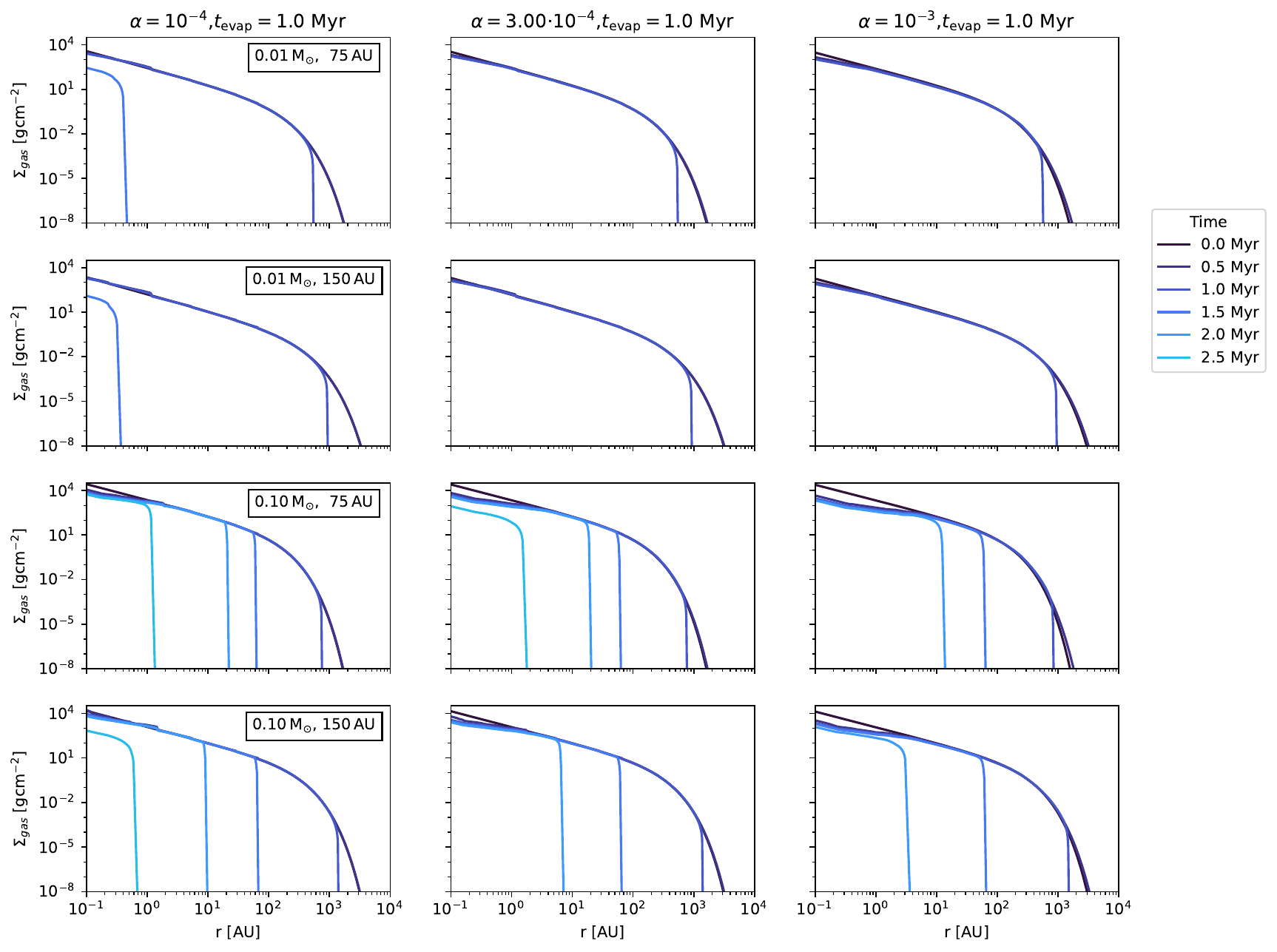}
   \caption{Same as Fig.~\ref{tevap0}, but the onset of the external photoevaporation time has been set 1.0 Myr. Note that the disc evolution time is about 1 Myr longer, because the later onset of external photoevaporation allows a longer disc life time.}
              \label{tevap1}%
    \end{figure*}

  \begin{figure*}[h]
   \centering
   \includegraphics[width=\textwidth]{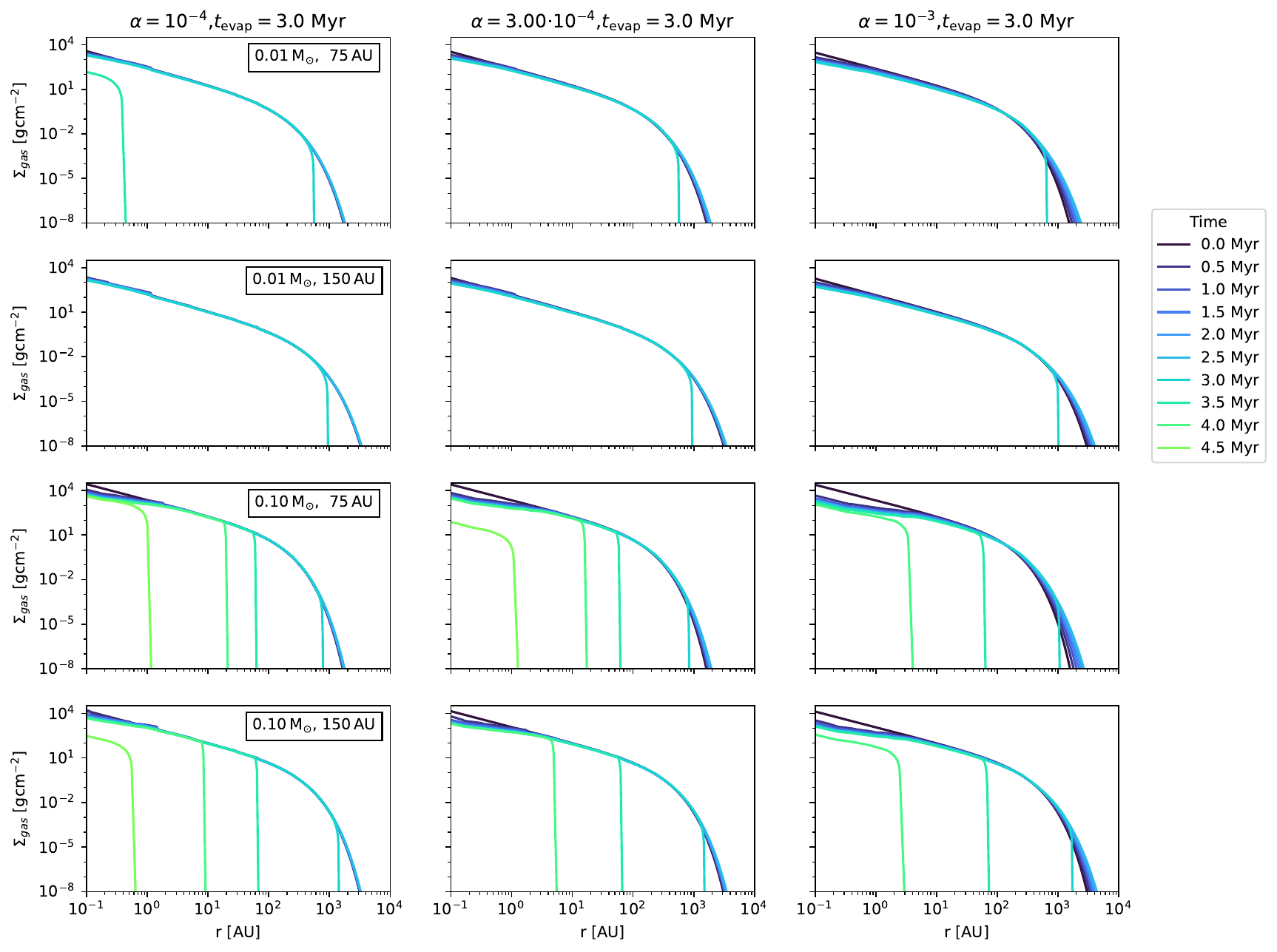}
   \caption{Same as Fig.~\ref{tevap0}, but with an onset of external photoevaporation after 3 Myr. Consequently, the disc survives longer compared to the previous models. }
              \label{tevap3}%
    \end{figure*}

\subsection{Gas surface density evolution}

In Figure~\ref{fig:sig_gas} (left panel), the results of the nominal simulation is illustrated, showing evolution of disc's gas surface density  over a span of 10 Myr, in the absence of external photoevaporation. The figure shows the natural dispersal mechanisms inherent to the disc's evolution, providing a fundamental baseline for our understanding of disc dynamics prior to considering the impact of external photoevaporative processes--the central focus of this paper.

Starting with scenarios of intense photoevaporation, exposing the disc to such conditions from as early as 0 Myr—without any protective shielding—generally results in significant truncation of the disc at early stages.. This observation is consistent with pioneering studies on shielding protoplanetary discs against external photoevaporation \citep{2022MNRAS.512.3788Q}. The extent of this impact, while dependent on specific disc parameters, is considerable in most cases. For instance, a disc with \(0.1 \, M_{\odot}\) and an initial radius of 75 AU within viscous environment of $\alpha = 10^{-4}$ is reduced to approximately 0.5 AU radius after just 1.5~Myr, indicating a 99 \% reduction in the disc radius and its capacity to harbor planetary building materials (Figure~\ref{tevap0}). The smaller mass discs of our sample \(0.01 \, M_{\odot}\) are basically dispersed immediately, because of the high evaporation rates and their low mass.

For higher viscosities, the picture is not changed significantly. In fact, the disc's life times are reduced, compared to the low viscosity simulations, because the higher viscosity allows a faster viscous evolution of the inner disc not touched upon by external photoevaporation.

Increasing the shielding time from photoevaporation reduces the destructive effects of photoevaporation on the disc profile. For example, with photoevaporation now beginning at 1 Myr, the disc with \(0.1 \, M_{\odot}\) can now survive until 2.5 Myr (third row and first column plot of Figure~\ref{tevap1}). With shielding increased further to 3 Myr, the disc can survive even favorably until 4 Myr (third row and first column plot of Figure~\ref{cotevap3}). Essentially, the disc's lifetime is increased by the delayed start of the external photoevaporation.

As external photoevaporation acts from the outside inwards, discs with a higher surface density harbor a longer lifetime. This can be seen by comparing the \(0.1 \, M_{\odot}\) discs with 75 and 150 AU characteristic radius. The smaller disc harbors a higher gas surface density and thus survives longer against external photoevaporation.

Therefore, the disc mass sets the lifetime against the external photoevaporation. Low mass disc with \(0.01 \, M_{\odot}\) are dispersed immediately, while higher mass discs can survive longer. This change in the disc's life time has important consequences for the chemical evolution of the inner disc regions, as well as on the formation of planets in the inner disc regions, which can be hindered if external photoevaporation is too efficient \citep{2023MNRAS.522.1939Q} and starts early. On the other hand, as giant planets form early \citep{2023A&A...679A..42S}, a delayed external photoevaporation by 1 Myr might not prevent the formation of planets, but might just reduce their final mass acquired by gas accretion.
\subsection{Chemical evolution of purely viscous disc}\label{app:chemistry}

\begin{figure*}[h]
    \centering
    \begin{minipage}{\textwidth}
        \subfigure{\includegraphics[width=0.33\textwidth]{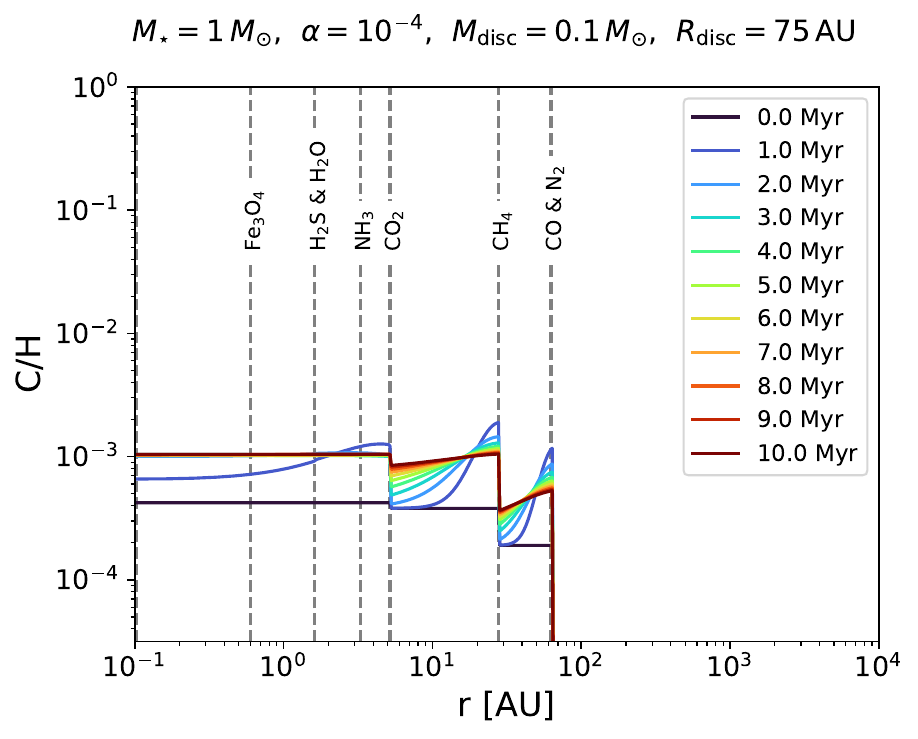}}
        \subfigure{\includegraphics[width=0.33\textwidth]{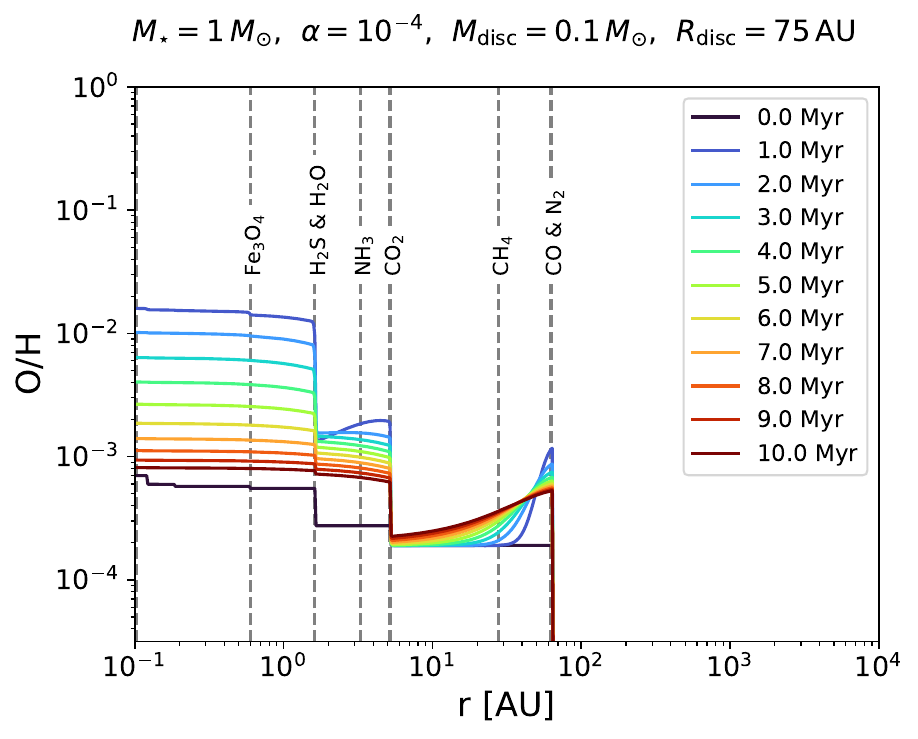}}
        \subfigure{\includegraphics[width=0.33\textwidth]{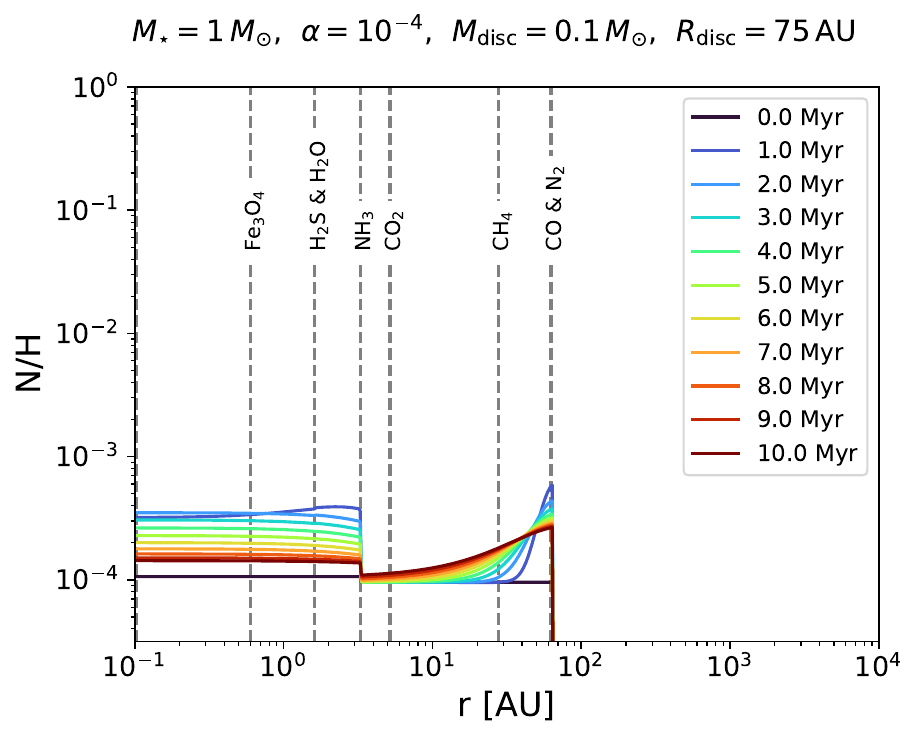}}
    \end{minipage}
    \caption{Different element ratios in the gas phase as a function of disc radius and time for a viscous disc without external photoevaporation (see Fig.~\ref{fig:sig_gas}) The element ratios feature disc with viscosity of 1e-4, disc mass of 0.1 $M_{\odot}$, and disc radius of 75~au. Left: C/H, middle: O/H, right: N/H. The color coding shows the time evolution.}
    \label{fig:CH_OH_NH_with_ph-nominal}
\end{figure*}

We show in Fig.~\ref{fig:CH_OH_NH_with_ph-nominal} the chemical evolution of a purely viscous disc corresponding to the disc shown in Fig.~\ref{fig:sig_gas}. In particular we show C/H, O/H and N/H as function of time. This allows a comparison of the chemical evolution to the disc subject to external photoevaporation.

As the pebbles drift inwards, they evaporate at their corresponding evaporation lines, enriching the disc with their vapour. The relative increase of the C/H, O/H and N/H to the initial value is governed by the abundance of the corresponding volatile molecules. For example, the increase in carbon at the CH$_4$ evaporation front is larger compared to the increase at the CO$_2$ evaporation front, because the total abundance of CH$_4$ is larger than the total abundance of CO$_2$ (see table~\ref{table:molecular_species}). For oxygen, obviously the jump at the water evaporation front is largest.

Once pebbles start to evaporate, the disc is enriched first closest to the evaporation front. With time the enrichment in vapour slowly moves inward with the gas, smoothing out the enrichment, resulting in a nearly horizontal enrichment of the disc interior to evaporation fronts in the inner disc. In the outer disc, this effect takes longer, as the distance between the evaporation fronts is larger. With larger viscosity, this effect happens faster, see also \citet{mahCloseinIceLines2023}.

\subsection{Pebble flux evolution}\label{app:flux}

\begin{figure}
    \centering
    \begin{minipage}{0.5\textwidth}
        \includegraphics[width=\textwidth]{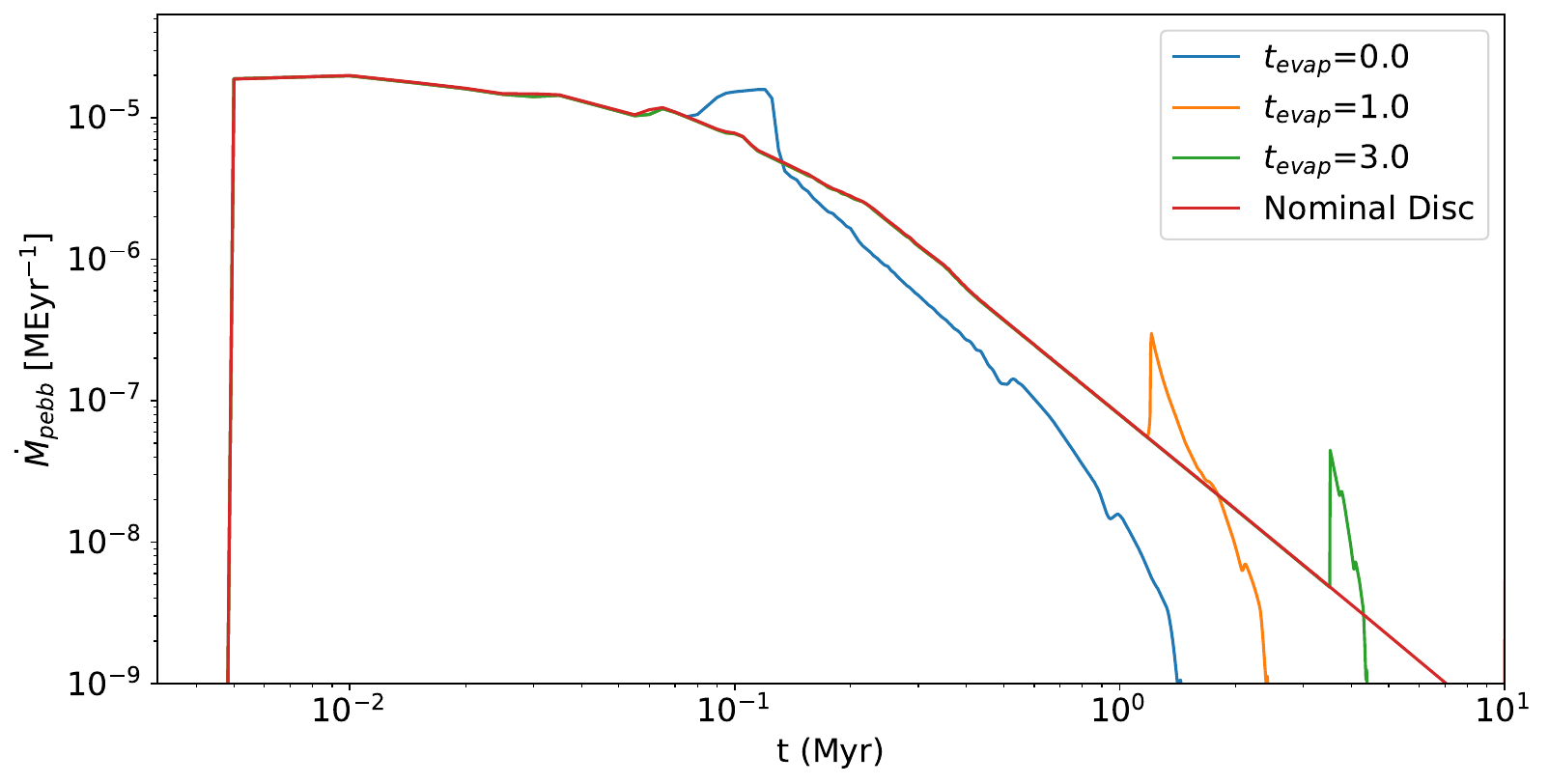}
    \end{minipage}
    \caption{Pebble flux at disc radial location of 0.5~AU for our nominal disc and discs with external photoevaporation beginning at 0, 1 and 3 Myr, respectively.}
    \label{fig:pebbflux_05au}
\end{figure}

We show in Fig.~\ref{fig:pebbflux_05au} the time evolution of the pebble flux in the inner disc (at r=0.5 AU) for a purely viscous disc and for discs where the external photoevaporation starts at 0, 1 and 3 Myr. The discs all feature an $\alpha$ viscosity parameter of $10^{-4}$ and have an initial disc radius of 75 AU with a mass of 0.1 solar masses, corresponding to the standard disc featured in Fig.~\ref{fig:sig_gas}.

Within the first few 10s of kyrs, there is no difference in the pebble flux between the different models. However, at around 100kyr, the pebble flux in the disc where external photoevaporation starts to act at t=0, increases compared to the purely viscous case. This is related to the fact that the outer disc loses gas density, resulting in an increase of the Stokes numbers of the particles in the outer disc which consequently drift inwards faster, enhancing the pebble flux. Consequently, this enlarged pebble flux depletes the outer disc of pebbles and the pebble flux is then reduced compared to the pure viscous case after around 150kyrs. The pebble flux then steadily declines as the outer disc starts to be evaporated away. At around 1.5 Myr, a steep decrease in the pebble flux is observed, due the fact that the whole disc is slowly removed by photoevaporation. 

The discs where external photoevaporation starts at 1 or 3 Myr, show a similar behavior. Shortly after the external photoevaporation starts, an increase in the pebble flux can be observed. This is again related to the larger Stokes number of the particles in the outer disc that then start to drift inwards effectively. But again, once these pebbles have drifted inwards, the pebble flux declines and rapidly falls off around 1.5 Myr after the external photoevaporation starts to be efficient. This is related - as described above - to the loss of material that rapidly drifted inwards due to mass loss of the disc.

Generally speaking, the pebble flux is the same in the purely viscous case and in discs with external photoevaporation, until external photoevaporation becomes efficient. Consequently, as discussed in the main paper, the inner disc composition driven by inward drifting and evaporating pebbles, is similar in these cases. On the other hand, the more rapid inward flux of pebbles might be hindering the formation of planets, because a larger pebble flux does not linearly translate into a faster growth of planets, especially when the planetary embryos are still small (e.g. \citealt{2014A&A...572A.107L, 2015A&A...582A.112B, 2023MNRAS.522.1939Q}. However, if external photoevaporation just starts after 1 Myr of evolution, the planets could have already formed \citep{2023A&A...679A..42S}, where we expect that external photoevaporation would just set the final mass of gas giants rather than preventing their formation in general.

\subsection{Model Parameters}

\begin{table}[h]
    \centering
    \small
	\begin{tabular}{ccc}
		\hline
        Species (Y)	        & $T_{\text{cond}}$ [K] & $v_{\text{Y}}$                    \\ \hline \hline
        CO                  & $20$                  & $0.45 \: \times \: \text{C/H}$    \\
        N$_2$               & $20$                  & $0.45 \: \times \: \text{N/H}$     \\
        CH$_4$              & $30$                  & $0.45 \: \times \: \text{C/H}$    \\
        CO$_2$              & $70$                  & $0.1 \: \times \: \text{C/H}$     \\
        NH$_3$              & $90$                  & $0.1 \: \times \: \text{N/H}$     \\
        H$_2$S              & $150$	                & $0.1 \: \times \: \text{S/H}$     \\
        H$_2$O              & $150$	                & $\text{O/H} - \left( 3 \times \text{MgSiO}_3/\text{H} + 4 \times                                                        \text{Mg}_2\text{SiO}_4/\text{H} + \text{CO/H} + 2 \times                                                               \text{CO}_2/\text{H} + 3 \times \text{Fe}_2\text{O}_3/\text{H} \right.$ \\
                            &                       & $ \left. + \: 4 \times \text{Fe}_3\text{O}_4/\text{H} + \text{VO/H} +                                  \text{TiO/H} + 3 \times \text{Al}_2\text{O}_3 + 8 \times                                            \text{NaAlSi}_3\text{O}_8 + 8 \times \text{KAlSi}_3\text{O}_8 \right)$ \\
        Fe$_3$O$_4$         & $371$	                & $(1/6) \times (\text{Fe/H} - 0.9 \times \text{S/H})$      \\
        FeS                 & $704$	                & $0.9 \: \times \: \text{S/H}$     \\
        NaAlSi$_3$O$_8$     & $958$	                & Na/H                              \\
        KAlSi$_3$O$_8$      & $1006$                & K/H                               \\
        Mg$_2$SiO$_4$       & $1354$                & $\text{Mg/H} - (\text{Si/H} - 3 \times \text{K/H} - 3 \times \text{Na/H})$    \\
        Fe$_2$O$_3$         & $1357$                & $0.25 \times (\text{Fe/H} - 0.9 \times \text{S/H})$       \\
        VO                  & $1423$                & V/H                               \\
        MgSiO$_3$           & $1500$                & $\text{Mg/H} - 2 \times (\text{Mg/H} - (\text{Si/H} - 3 \times \text{K/H} - 3 \times \text{Na/H}))$   \\
        Al$_2$O$_3$         & $1653$                & $0.5 \times (\text{Al/H} - (\text{K/H} + \text{Na/H}))$   \\
        TiO                 & $2000$                & Ti/H                              \\ \hline
	\end{tabular}
    \caption{Condensation temperatures and volume mixing ratios of the chemical species in our model. The condensation temperatures are taken from \cite{loddersSolarSystemAbundances2003}. We assume constant $T_{\text{cond}}$ as its dependence on pressure is marginal \citep{mousisDETERMINATIONMINIMUMMASSES2009}. For simplification, we treat condensation and sublimation temperature as the same. Note that for Fe$_2$O$_3$, the condensation temperature for pure iron is adopted \citep{loddersSolarSystemAbundances2003}. The $v_{\text{Y}}$'s (i.e., by number) are adopted for the species as a function of disc elemental abundances \citep[e.g.,][]{madhusudhanExoplanetaryAtmospheres2014}. They are expanded on the different mixing ratios from \cite{bitschInfluenceSubSupersolar2020}.}
    \label{table:molecular_species}
\end{table}

\begin{table}
    \centering
	\begin{tabular}{cc}
		\hline
        Species (X)	& Abundance             \\ \hline \hline
        He/H        & $0.085$               \\
        O/H         & $4.90 \cdot 10^{-4}$  \\
        C/H         & $2.69 \cdot 10^{-4}$	\\
        N/H         & $6.76 \cdot 10^{-5}$	\\
        Mg/H        & $3.98 \cdot 10^{-5}$	\\
        Si/H        & $3.24 \cdot 10^{-5}$	\\
        Fe/H        & $3.16 \cdot 10^{-5}$	\\
        S/H         & $1.32 \cdot 10^{-5}$	\\
        Al/H        & $2.82 \cdot 10^{-6}$	\\
        Na/H        & $1.74 \cdot 10^{-6}$	\\
        K/H         & $1.07 \cdot 10^{-7}$	\\
        Ti/H        & $8.91 \cdot 10^{-8}$	\\
        V/H         & $8.59 \cdot 10^{-9}$	\\ \hline
	\end{tabular}
    \caption{Number ratios of the elements used in our model, corresponding to the abundance of element X relative to hydrogen in the solar photosphere \citep{asplundChemicalCompositionSun2009}.}
    \label{table:initial_abundances}
\end{table}

We show in this section the input parameters for our chemical composition model. In table~\ref{table:molecular_species} we show the partitioning of the different elements into molecules and solids, while table~\ref{table:initial_abundances} shows the initial elemental abundances used in our work.

\end{document}